\documentclass[journal]{IEEEtran}

\usepackage{cite}

\usepackage[colorinlistoftodos,prependcaption,textsize=small]{todonotes}
\usepackage{color}
\usepackage{amsmath,amssymb,graphicx,booktabs,url}
\usepackage[utf8]{inputenc}

\begin{document}
	\title{Multi-talker Speech Separation with Utterance-level Permutation Invariant Training of Deep Recurrent Neural Networks}
	
	
	\author{\IEEEauthorblockN{Morten Kolbæk,~\IEEEmembership{Student Member,~IEEE},
			Dong Yu,~\IEEEmembership{Senior Member,~IEEE}, \\
			Zheng-Hua Tan, ~\IEEEmembership{Senior Member,~IEEE}, and
			Jesper Jensen.} 
 
 
	\thanks{M. Kolbæk and Z.-H. Tan are with the Department of Electronic Systems, Aalborg University,
		Aalborg 9220, Denmark (e-mail: mok@es.aau.dk; zt@es.aau.dk).}%
	
	\thanks{D. Yu, corresponding author, is with Tencent AI Lab, Bellevue WA, USA. Part of work was done while he was at Microsoft Research (e-mail: dongyu@ieee.org).}
	 
	\thanks{J. Jensen is with the Department of Electronic Systems, Aalborg University, Aalborg 9220, Denmark, and also with Oticon A/S, Smørum 2765, Denmark (e-mail: jje@es.aau.dk; jesj@oticon.com).}%
	
		
	}%

	\IEEEtitleabstractindextext{%
	\begin{abstract}
		In this paper we propose the utterance-level Permutation Invariant Training\;(uPIT) technique. 
		uPIT is a practically applicable, end-to-end, deep learning based solution for speaker independent multi-talker speech separation. 
			Specifically, uPIT extends the recently proposed Permutation Invariant Training\;(PIT) technique with an utterance-level cost function, hence eliminating the need for solving an additional permutation problem during inference, which is otherwise required by frame-level PIT. 
			We achieve this using Recurrent Neural Networks\;(RNNs) that, during training, minimize the utterance-level separation error, hence forcing separated frames belonging to the same speaker to be aligned to the same output stream.    
			In practice, this allows RNNs, trained with uPIT, to separate multi-talker mixed speech without any prior knowledge of signal duration, number of speakers, speaker identity or gender.

			We evaluated uPIT on the WSJ0 and Danish two- and three-talker mixed-speech separation tasks and found that uPIT outperforms techniques based on Non-negative Matrix Factorization\;(NMF) and Computational Auditory Scene Analysis\;(CASA), and compares favorably with Deep Clustering\;(DPCL) and the Deep Attractor Network\;(DANet). Furthermore, we found that models trained with uPIT generalize well to unseen speakers and languages.
			Finally, we found that a single model, trained with uPIT, can handle both two-speaker, and three-speaker speech mixtures.	
	\end{abstract}

	}

\maketitle

\IEEEdisplaynontitleabstractindextext

%
\IEEEpeerreviewmaketitle

\section{Introduction}\label{sec:intro}
\IEEEPARstart{H}{aving} a conversation in a complex acoustic environment, with multiple noise sources and competing background speakers, is a task humans are remarkably good at \cite{haykin_cocktail_2005,bronkhorst_cocktail_2000}. 
The problem that humans solve when they focus their auditory attention towards one audio signal in a complex mixture of signals is commonly known as the cocktail party problem \cite{haykin_cocktail_2005,bronkhorst_cocktail_2000}.
Despite intense research for more than half a century, a general machine based solution to the cocktail party problem is yet to be discovered \cite{cherry_experiments_1953,bronkhorst_cocktail_2000,haykin_cocktail_2005,cooke_monaural_2010}. 
A machine solution to the cocktail party problem is highly desirable for a vast range of applications. These include automatic meeting transcription, automatic captioning for audio/video recordings (e.g., YouTube), multi-party human-machine interaction (e.g., in the world of Internet of things (IoT)), and advanced hearing aids, where overlapping speech is commonly encountered.

Since the cocktail party problem was initially formalized \cite{cherry_experiments_1953}, a large number of potential solutions have been proposed \cite{divenyi_speech_2005}, and the most popular techniques originate from the field of Computational Auditory Scene Analysis\;(CASA) \cite{ellis_prediction-driven_1996,cooke_modelling_2005,wang_computational_2006,shao_model-based_2006,hu_unsupervised_2013}. 
In CASA, different segmentation and grouping rules are used to group Time-Frequency\;(T-F) units that are believed to  belong to the same speaker. The rules are typically hand-engineered and based on heuristics such as pitch trajectory, common onset/offset, periodicity, etc. The grouped T-F units are then used to extract a particular speaker from the mixture signal.
Another popular technique for multi-talker speech separation is Non-negative Matrix Factorization\;(NMF) \cite{schmidt_single-channel_2006,smaragdis_convolutive_2007,roux_sparse_2015,lee_algorithms_2000}. The NMF technique uses non-negative dictionaries to decompose the spectrogram of the mixture signal into speaker specific activations, and from these activations an isolated target signal can be approximated using the dictionaries.
For multi-talker speech separation, both CASA and NMF have led to limited success \cite{cooke_monaural_2010,divenyi_speech_2005} and the most successful techniques, before the deep learning era, are based on probabilistic models \cite{kristjansson_super-human_2006,virtanen_speech_2006,stark_source-filter-based_2011}, such as factorial GMM-HMM \cite{ghahramani_factorial_1997}, that model the temporal dynamics and the complex interactions of the target and competing speech signals. Unfortunately, these models assume and only work under closed-set speaker conditions, i.e. the identity of the speakers must be known \emph{a priori}. 

More recently, a large number of techniques based on deep learning\cite{goodfellow_deep_2016} have been proposed, especially for Automatic Speech Recognition\;(ASR) \cite{yu_roles_2010,dahl_context-dependent_2012,seide_conversational_2011,hinton_deep_2012,xiong_achieving_2016,saon_english_2017}, and speech enhancement  \cite{wang_towards_2013,wang_training_2014,xu_experimental_2014,weninger_speech_2015,huang_joint_2015,chen_large-scale_2016,kolbaek_speech_2017,du_speech_2014,goehring_speech_2017}. Deep learning has also been applied in the context of multi-talker speech separation (e.g., \cite{huang_joint_2015}), although successful work has, similarly to NMF and CASA, mainly been reported for closed-set speaker conditions.      

The limited success in deep learning based speaker independent multi-talker speech separation is partly due to the label permutation problem (which will be described in detail in Sec.~\ref{sec:PIT}).
To the authors knowledge only four deep learning based works \cite{weng_deep_2015,hershey_deep_2016,chen_deep_2017,yu_permutation_2017} exist, that have tried to address and solve the harder speaker independent multi-talker speech separation task. 

In Weng \emph{et al.} \cite{weng_deep_2015}, which proposed the best performing system in the 2006 monaural speech separation and recognition challenge \cite{cooke_monaural_2010}, the instantaneous energy was used to determine the training label assignment, which alleviated the label permutation problem and allowed separation of unknown speakers. Although this approach works well for two-speaker mixtures, it is hard to scale up to mixtures of three or more speakers.   

Hershey \emph{et al.} \cite{hershey_deep_2016} have made significant progress with their Deep Clustering\;(DPCL) technique. In their work, a deep Recurrent Neural Network\;(RNN) is used to project the speech mixture into an embedding space, where T-F units belonging to the same speaker form a cluster. In this embedding space a clustering algorithm (e.g. K-means) is used to identify the clusters. Finally, T-F units belonging to the same clusters are grouped together and a binary mask is constructed and used to separate the speakers from the mixture signal. 
To further improve the model \cite{isik_single-channel_2016}, another RNN is stacked on top of the first DPCL RNN to estimate continuous masks for each target speaker. 
Although DPCL show good performance, the technique is potentially limited because the objective function is based on the affinity between the sources in the embedding space, instead of the separated signals themselves. That is, low proximity in the embedding space does not necessarily imply perfect separation of the sources in the signal space.    

Chen \emph{et al.} \cite{chen_deep_2017,chen_single_2017} proposed a related technique called Deep Attractor Network\;(DANet). Following DPCL, the DANet approach also learns a high-dimensional embedding of the mixture signals. Different from DPCL, however, it creates attractor points (cluster centers) in the embedding space, which attract the T-F units corresponding to each target speaker. The training is conducted in a way similar to the Expectation Maximization\;(EM) principle. The main disadvantage of DANet over DPCL is the added complexity associated with estimating attractor points during inference.

Recently, we proposed the Permutation Invariant Training\;(PIT) technique%
\footnote{In \cite{hershey_deep_2016}, a related permutation free technique, which is similar to PIT for exactly two-speakers, was evaluated with negative results and conclusion.}
\cite{yu_permutation_2017} for attacking the speaker independent multi-talker speech separation problem and showed that PIT effectively solves the label permutation problem.
However, although PIT solves the label permutation problem at training time, PIT does not effectively solve the permutation problem during inference, where the permutation of the separated signals at the frame-level is unknown. We denote the challenge of identifying this frame-level permutation, as the \emph{speaker tracing problem}. 

In this paper, we extend PIT and propose an utterance-level Permutation Invariant Training\;(uPIT) technique, which is a practically applicable, end-to-end, deep learning based solution for speaker independent multi-talker speech separation.
Specifically, uPIT extends the frame-level PIT technique \cite{yu_permutation_2017} with an utterance-level training criterion that effectively eliminates the need for additional speaker tracing or very large input/output contexts, which is otherwise required by the original PIT \cite{yu_permutation_2017}. 
We achieve this using deep Long Short-Term Memory\;(LSTM) RNNs \cite{hochreiter_long_1997} that, during training, minimize the utterance-level separation error, hence forcing separated frames belonging to the same speaker to be aligned to the same output stream. 
This is unlike other techniques, such as DPCL and DANet, that require a distinct clustering step to separate speakers during inference. Furthermore, the computational cost associated with the uPIT training criterion is negligible compared to the computations required by the RNN during training and is zero during inference. 
We evaluated uPIT on the WSJ0 and Danish two- and three-talker mixed-speech separation tasks and found that uPIT outperforms techniques based on NMF and CASA, and compares favorably with DPCL and DANet. 
Furthermore, we show that models trained with uPIT generalize well to unseen speakers and languages, and finally, we found that a single model trained with uPIT can separate both two-speaker, and three-speaker speech mixtures.

The rest of the paper is organized as follows. In Sec.~\ref{sec:problem} we describe the monaural speech separation problem. In Sec.~\ref{sec:mask} we extend popular optimization criteria used in separating single-talker speech from noises, to multi-talker speech separation tasks. 
In Sec.~\ref{sec:PIT} we discuss the label permutation problem and present the PIT framework. 
In Sec.~\ref{sec:integrated} we introduce uPIT and show how an utterance-level permutation criterion can be combined with PIT. We report series of experimental results in Sec.~\ref{sec:exp} and conclude the paper in Sec.~\ref{sec:conclusion}.

	\section{Monaural speech separation}\label{sec:problem}
	
	The goal of monaural speech separation is to estimate the individual source signals ${x}_s[n], s=1,\cdots,S$ in a linearly mixed single-microphone signal  
	\begin{equation}
		y[n]=\sum_{s=1}^{S} {x}_s[n],
	\end{equation}	
	based on the observed signal ${y}[n]$ only. 
	In real situations, the received signals may be reverberated, i.e., the underlying clean signals are filtered before being observed in the mixture. In this condition, we aim at recovering the reverberated source signals ${x}_s[n]$, i.e., we are not targeting the dereverberated signals.
	
	The separation is usually carried out in the T-F domain, in which the task can be cast as recovering the Short-Time discrete Fourier Transformation\;(STFT) of the source signals  ${X}_s(t,f)$ for each time frame $t$ and frequency bin $f$, given the mixed speech 
	\begin{equation}
	\begin{split}
	{Y}(t,f) =\sum_{n=0}^{N-1} {y}[n+tL]w[n]\exp(-j2{\pi}nf/N),
	\end{split}
	\end{equation}
	where $w[n]$ is the analysis window of length $N$, the signal is shifted by an amount of $L$ samples for each time frame $t=0,\cdots,T-1$, and each frequency bin $f=0,\cdots,N-1$ is corresponding to a frequency of $(f/N)f_s$ [Hz] when the sampling rate is $f_s$ [Hz]. 

	From the estimated STFT ${\hat{X}}_s(t,f)$ of each source signal, an inverse Discrete Fourier Transform\;(DFT) 
	\begin{equation}
		{\hat{x}}_{s,t}[n] =\frac{1} {N} \sum_{f=0}^{N-1} \hat{X}_s(t,f)\exp(j2{\pi}nf/N)    
	\end{equation}
	can be used to construct estimated time-domain frames, and the overlap-add operation
	\begin{equation}
		{\hat{x}_s}[n] =\sum_{t=0}^{T-1} v[n-tL] \hat{x}_{s,t}[n-tL]              
	\end{equation}
	can be used to reconstruct the estimate $\hat{x}_s[n]$ of the original signal, where $v[n]$ is the synthesis window.

	In a typical setup, however, only the STFT magnitude spectrum $A_s(t,f) \triangleq |{X}_s(t,f)|$ is estimated from the mixture during the separation process, and the phase of the mixed speech is used directly, when recovering the time domain waveforms of the separated sources. This is because phase estimation is still an open problem in the speech separation setup \cite{williamson_complex_2016,erdogan_deep_2017}.
	Obviously, given only the magnitude of the mixed spectrum, $R(t,f)\triangleq |{Y}(t,f)|$, the problem of recovering $A_s(t,f)$ is under-determined, as there are an infinite number of possible $A_s(t,f)$, $s=1, \dots ,S$ combinations that lead to the same $R(t,f)$. To overcome this problem, a supervised learning system has to learn from some training set $\mathbb{S}$ that contains corresponding observations of $R(t,f)$ and $A_s(t,f)$, $s=1, \dots ,S$.

    Let $\mathbf{a}_{s,i} = \left[ {{A_s}(i,1)} ,\; {{A_s}(i,2)} ,\; \dots\; {{A_s}(i,\frac{N}{2}+1)} \right]^T \in \mathbb{R}^{\frac{N}{2}+1}$ denote the single-sided magnitude spectrum for source $s$ at frame $i$.
    Furthermore, let $\mathbf{A}_s \in \mathbb{R}^{\left(\frac{N}{2}+1\right) \times T}$ be the single-sided magnitude spectrogram for source $s$ and all frames $i=1, \dots ,T$, defined as $\mathbf{A}_s = \left[\mathbf{a}_{s,1} ,\; \mathbf{a}_{s,2} ,\; \dots\;, \mathbf{a}_{s,T}\right]$.  
    Similarly, let $\mathbf{r}_{i} = \left[ {{R}(i,1)} ,\; {{R}(i,2)} ,\; \dots\; {{R}(i,\frac{N}{2}+1)} \right]^T$ be the single-sided magnitude spectrum of the observed signal at frame $i$ and let $\mathbf{R} = \left[\mathbf{r}_{1} ,\; \mathbf{r}_{2} ,\; \dots\;, \mathbf{r}_{T}\right] \in \mathbb{R}^{\left(\frac{N}{2}+1\right) \times T}$ be the single-sided magnitude spectrogram for all frames $i=1, \dots ,T$.

	Furthermore, let us denote a supervector $\mathbf{z}_{i} = \left[ \mathbf{a}_{1,i}^T \; \mathbf{a}_{2,i}^T \; \dots \; \mathbf{a}_{S,i}^T \right]^T \in \mathbb{R}^{S\left(\frac{N}{2}+1\right)}$, consisting of the stacked source magnitude spectra for each source $s = 1, \dots, S$ at frame $i$ and let $\mathbf{Z} = \left[\mathbf{z}_{1} ,\; \mathbf{z}_{2} ,\; \dots\;, \mathbf{z}_{T}\right] \in \mathbb{R}^{S\left(\frac{N}{2}+1\right) \times T}$ denote the matrix of all $T$ supervectors.
	Finally, let $\mathbf{y}_{i} = \left[ {{Y}(i,1)} ,\; {{Y}(i,2)} ,\; \dots\; {{Y}(i,\frac{N}{2}+1)} \right]^T \in \mathbb{C}^{\frac{N}{2}+1}$ be the single-sided STFT of the observed mixture signal at frame $i$ and $\mathbf{Y} = \left[\mathbf{y}_{1} ,\; \mathbf{y}_{2} ,\; \dots\;, \mathbf{y}_{T}\right] \in \mathbb{C}^{\left(\frac{N}{2}+1\right) \times T} $ be the STFT of the mixture signal for all $T$ frames.  

	Our objective is then to train a deep learning model $g(\cdot)$, parameterized by a parameter set $\mathbf{\Phi}$, such that $g\left(d\left(\mathbf{Y}\right);\mathbf{\Phi}\right)=\mathbf{Z}$, where $d(\mathbf{Y})$ is some feature representation of the mixture signal: In a particularly simple situation, $d(\mathbf{Y}) = \mathbf{R}$, i.e., the feature representation is simply the magnitude spectrum of the observed mixture signal.    
	
	It is possible to directly estimate the magnitude spectra $\mathbf{Z}$ of all sources using a deep learning model. However, it is well-known (e.g., \cite{wang_training_2014, erdogan_deep_2017}), that better results can be achieved if, instead of estimating $\mathbf{Z}$ directly, we first estimate a set of masks ${M_s}(t,f)$, $s=1, \dots ,S$. 

	Let $\mathbf{m}_{s,i} = \left[ {{M_s}(i,1)} \;,\; {{M_s}(i,2)} \;,\; \dots\; {{M_s}(i,\frac{N}{2}+1)} \right]^T \in \mathbb{R}^{\frac{N}{2}+1}$ be the ideal mask (to be defined in detail in Sec.\;\ref{sec:mask}) for speaker $s$ at frame $i$, and let $\mathbf{M}_s = \left[\mathbf{m}_{s,1} ,\; \mathbf{m}_{s,2} ,\; \dots\;, \mathbf{m}_{s,T}\right] \in \mathbb{R}^{\left(\frac{N}{2}+1\right) \times T}$ be the ideal mask for all $T$ frames, 	
	such that $\mathbf{A}_s = \mathbf{M}_s \circ \mathbf{R} $, where $\circ$ is the Hadamard product, i.e. element-wise product of two operands.
	Furthermore, let us introduce the mask supervector $\mathbf{u}_{i} = \left[ \mathbf{m}_{1,i}^T \; \mathbf{m}_{2,i}^T \; \dots \; \mathbf{m}_{S,i}^T \right]^T \in \mathbb{R}^{S\left(\frac{N}{2}+1\right)}$ and the corresponding mask matrix $\mathbf{U} = \left[\mathbf{u}_{1} ,\; \mathbf{u}_{2} ,\; \dots\;, \mathbf{u}_{T}\right] \in \mathbb{R}^{S\left(\frac{N}{2}+1\right) \times T}$.
    Our goal is then to find an estimate $\hat{\mathbf{U}}$ of $\mathbf{U}$, using a deep learning model,
 	$h\left(\mathbf{R};\mathbf{\Phi} \right)=\hat{\mathbf{U}}$. 
 	Since,  $\hat{\mathbf{U}} = \left[\hat{\mathbf{u}}_{1} ,\; \hat{\mathbf{u}}_{2} ,\; \dots\;, \hat{\mathbf{u}}_{T}\right]$ and $\hat{\mathbf{u}}_{i} = \left[ \hat{\mathbf{m}}_{1,i}^T \; \hat{\mathbf{m}}_{2,i}^T \; \dots \; \hat{\mathbf{m}}_{S,i}^T \right]^T$, the model output is easily divided into output streams corresponding to the estimated masks for each speaker $\hat{\mathbf{m}}_{s,i}$, and their resulting magnitudes are estimated as $\hat{\mathbf{a}}_{s,i} = \hat{\mathbf{m}}_{s,i} \circ \mathbf{r}_{i}$.  
 	The estimated time-domain signal for speaker $s$ is then computed as the inverse DFT of $\hat{\mathbf{a}}_{s,i}$ using the phase of the mixture signal $\mathbf{y}_i$.

	\section{Masks and training criteria}\label{sec:mask}
	Since masks are to be estimated as an intermediate step towards estimating magnitude spectra of source signals, we extend in the following three popular masks defined for separating single-talker speech from noises to the multi-talker speech separation task at hand.

	\subsection{Ideal Ratio Mask}
	The Ideal Ratio Mask\;(IRM) \cite{wang_training_2014} for each source is defined as 
	\begin{equation}
		{M}_s^{irm}(t,f) =\frac {|{X}_s(t,f)|} { \sum_{s=1}^{S} |{X}_s(t,f)|}  .           
	\end{equation}
	When the phase of $\mathbf{Y}$ is used for reconstruction, the IRM achieves the highest Signal to Distortion Ratio\;(SDR) \cite{vincent_performance_2006}, when all sources have the same phase, (which is an invalid assumption in general). 
	IRMs are constrained to $0 \leq {{M}}_s^{irm}(t,f) \leq 1$ and $\sum_{s=1}^S {{M}}_s^{irm}(t,f) = 1$  for all T-F units. This constraint can easily be satisfied using the softmax activation function. 

	Since $\mathbf{Y}$ is the only observed signal in practice and $\sum_{s=1}^{S} |{X}_s(t,f)|$ is unknown during separation, the IRM is not a desirable target for the problem at hand. Nevertheless, we report IRM results as an upper performance bound since the IRM is a commonly used training target for deep learning based monaural speech separation \cite{chen_large-scale_2016,kolbaek_speech_2017}.

	\subsection{Ideal Amplitude Mask}
	Another applicable mask is the Ideal Amplitude Mask\;(IAM) (known as FFT-mask in \cite{wang_training_2014}), or simply Amplitude Mask\;(AM), when estimated by a deep learning model. The IAM is defined as 
	\begin{equation}
		{M}_s^{iam}(t,f) =\frac {|{X}_s(t,f)|} {|{Y}(t,f)|}.             
	\end{equation}
	
	Through IAMs we can construct the exact $|{X}_s(t,f)|$ given the magnitude spectra of the mixed speech $|{Y}(t,f)|$. If the phase of each source equals the phase of the mixed speech, the IAM achieves the highest SDR. Unfortunately, as with the IRM, this assumption is not satisfied in most cases. IAMs satisfy the constraint that $0 \leq {{M}}_s^{iam}(t,f) \leq \infty$, although we found empirically that the majority of the T-F units are in the range of $0 \leq {{M}}_s^{iam}(t,f) \leq 1$. For this reason, softmax, sigmoid and ReLU are all possible output activation functions for estimating IAMs.

    \subsection{Ideal Phase Sensitive Mask}
	Both IRM and IAM do not consider phase differences between source signals and the mixture. This leads to sub-optimal results, when the phase of the mixture is used for reconstruction.  
	The Ideal Phase Sensitive Mask\;(IPSM) \cite{erdogan_phase-sensitive_2015,erdogan_deep_2017}
	\begin{equation}
		{M}_s^{ipsm}(t,f) =\frac {|{X}_s(t,f)|\cos(\theta_y(t,f)-\theta_s(t,f))} {|{Y}(t,f)|},             
	\end{equation}
	however, takes phase differences into consideration, where $\theta_y$ and $\theta_s$ are the phases of mixed speech ${Y}(t,f)$ and source ${X}_s(t,f)$, respectively. Due to the phase-correcting term, the IPSM sums to one, i.e. $\sum_{s=1}^S {{M}}_s^{ipsm}(t,f) = 1$. Note that since $|\cos(\cdot)| \leq 1$ the IPSM is smaller than the IAM, especially when the phase difference between the mixed speech and the source is large. 
	
	Even-though the IPSM in theory is unbounded, we found empirically that the majority of the IPSM is in the range of $0 \leq {M}_s^{ipsm}(t,f) \leq 1$. Actually, in our study we have found that approximately $20\%$ of IPSMs are negative. However, those negative IPSMs usually are very close to zero. To account for this observation, we propose the Ideal Non-negative PSM\;(INPSM), which is defined as 
	\begin{equation}
		{M}_s^{inpsm}(t,f) =max(0, {M}_s^{ipsm}(t,f)).             
	\end{equation}
	For estimating the IPSM and INPSM, Softmax, Sigmoid, tanh, and ReLU are all possible activation functions, and similarly to the IAM, when the IPSM is estimated by a deep learning model we refer to it as PSM.

	\subsection{Training Criterion}
	Since we first estimate masks, through which the magnitude spectrum of each source can be estimated, the model parameters can be optimized to minimize the Mean Squared Error\;(MSE) between the estimated mask $\hat{M}_s$ and one of the target masks defined above as
	\begin{equation}
		J_m=\frac{1}{B}\sum_{s=1}^S \|\hat{\mathbf{M}}_s - \mathbf{M}_s\|_F^2,
	\end{equation}
	where $B=T \times N \times S$ is the total number of T-F units over all sources and $\|\cdot\|_F$ is the Frobenius norm. This approach comes with two problems. First, in silence segments, $|{X}_s(t,f)|=0$ and $|{Y}(t,f)|=0$, so that the target masks ${M}_s(t,f)$ are not well defined. Second, what we really care about is the error between the reconstructed source signal and the true source signal. 
	
	To overcome these limitations, recent works \cite{wang_training_2014} directly minimize the MSE
	\begin{equation}
	\begin{split}
	J_a & =\frac{1}{B}\sum_{s=1}^S \| \hat{\mathbf{A}}_s - \mathbf{A}_s \|_F^2  \\
	& =\frac{1}{B}\sum_{s=1}^S \| \hat{\mathbf{M}}_s \circ \mathbf{R} - \mathbf{A}_s \|_F^2 
	\end{split}
	\label{eq:am}
	\end{equation}
	between the estimated magnitude, i.e. $\hat{\mathbf{A}}_s = \hat{\mathbf{M}}_s \circ \mathbf{R}$ and the true magnitude $\mathbf{A}_s$. Note that in silence segments $A_s(t,f)=0$ and $R(t,f)=0$,  so the accuracy of mask estimation does not affect the training criterion for those segments. 
	Furthermore, using Eq.\;\eqref{eq:am} the IAM is estimated as an intermediate step. 
	
	When the PSM is used, the cost function becomes 
	\begin{equation}
		\begin{split}
			J_{psm}  = \frac{1}{B}\sum_{s=1}^S \|\hat{\mathbf{M}}_s  \circ \mathbf{R} - \mathbf{A}_s  \circ \cos(\mathbf{\theta}_y - \mathbf{\theta}_s)\|_F^2. 
		\end{split}
		\label{eq:psm}
	\end{equation}
	
	In other words, using PSMs is as easy as replacing the original training targets with the phase discounted targets. 
	Furthermore, when Eq.\;\eqref{eq:psm} is used as a cost function, the IPSM is the upper bound achievable on the task \cite{erdogan_deep_2017}.

	\section{Permutation Invariant Training}\label{sec:PIT}

	\subsection{Conventional Multi-Talker Separation}
\begin{figure}[ht]
		\centering
		\includegraphics[trim={5mm 5mm 5mm 5mm},clip,width=1.0\linewidth]{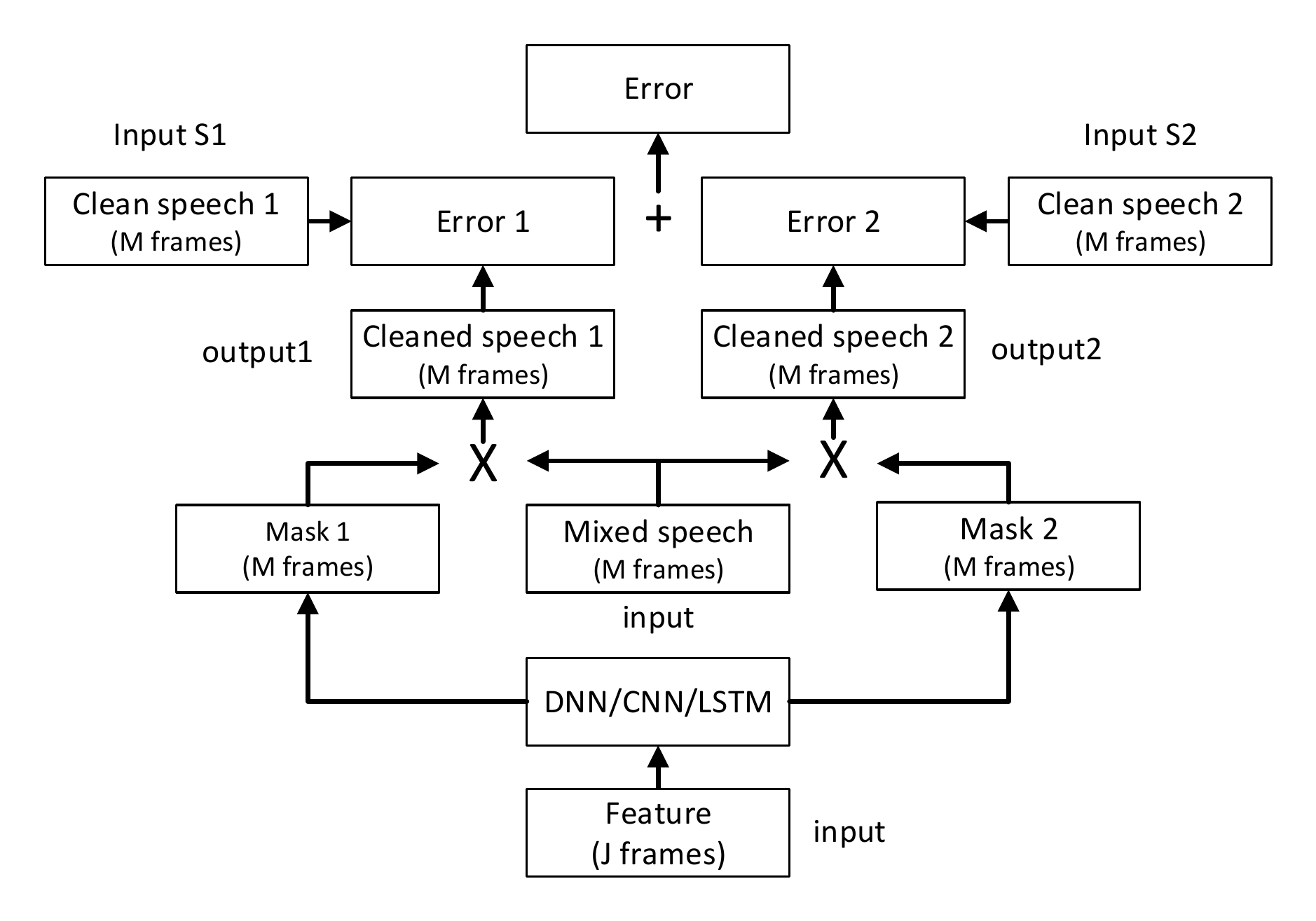}
		\caption{The conventional two-talker speech separation model.}\label{fig:conventional}
	\end{figure}

	A natural, and commonly used, approach for deep learning based speech separation is to cast the problem as a multi-class \cite{huang_joint_2015,tu_deep_2014,weng_deep_2015} regression problem as depicted in Fig.~\ref{fig:conventional}.  
	
	For this conventional two-talker separation model, $J$ frames of feature vectors of the mixed signal $\mathbf{Y}$ are used as the input to some deep learning model e.g. a feed-forward Deep Neural Network\,(DNN), Convolutional Neural Network\;(CNN), or LSTM RNN, to generate $M$ frames of masks for each talker.  
	Specifically, if $M=1$, the output of the model can be described by the vector $\hat{\mathbf{u}}_{i} = \left[ \hat{\mathbf{m}}_{1,i}^T \; \hat{\mathbf{m}}_{2,i}^T \right]^T$ and the  sources are separated as $\hat{\mathbf{a}}_{1,i} = \hat{\mathbf{m}}_{1,i} \circ \mathbf{r}_{i}$ and $\hat{\mathbf{a}}_{2,i} = \hat{\mathbf{m}}_{2,i} \circ \mathbf{r}_{i}$, for sources $s = 1,2$, respectively.

	\subsection{The Label Permutation Problem}
	During training, the error (e.g. using Eq.\;\eqref{eq:psm}) between the clean magnitude spectra $\mathbf{a}_{1,i}$ and $\mathbf{a}_{2,i}$ and their estimated counterparts $\hat{\mathbf{a}}_{1,i}$ and $\hat{\mathbf{a}}_{2,i}$ needs to be computed.
    However, since the model estimates the masks $\hat{\mathbf{m}}_{1,i}$ and $\hat{\mathbf{m}}_{2,i}$ simultaneously, and they depend on the same input mixture, it is unknown in advance whether the resulting output vector $\hat{\mathbf{u}}_{i}$ is ordered as $\hat{\mathbf{u}}_{i} = \left[ \hat{\mathbf{m}}_{1,i}^T \; \hat{\mathbf{m}}_{2,i}^T \right]^T$ or $\hat{\mathbf{u}}_{i} = \left[ \hat{\mathbf{m}}_{2,i}^T \; \hat{\mathbf{m}}_{1,i}^T \right]^T$. That is, the permutation of the output masks is unknown. 
    
    A na\"{i}ve approach to train a deep learning separation model, without exact knowledge about the permutation of the output masks, is to use a constant permutation as illustrated by Fig.~\ref{fig:conventional}. 
    Although such a training approach works for simple cases e.g. female speakers mixed with male speakers, in which case \emph{a priori} convention can be made that e.g. the first output stream contains the female speaker, while the second output stream is paired with the male speaker, the training fails if the training set consists of many utterances spoken by many speakers of both genders.

	This problem is referred to as the label permutation (or ambiguity) problem in \cite{weng_deep_2015,hershey_deep_2016}. Due to this problem, prior arts perform poorly on speaker independent multi-talker speech separation.

	\subsection{Permutation Invariant Training}
\begin{figure}[ht]
		\centering
		\includegraphics[trim={5mm 5mm 5mm 5mm},clip,width=1\linewidth]{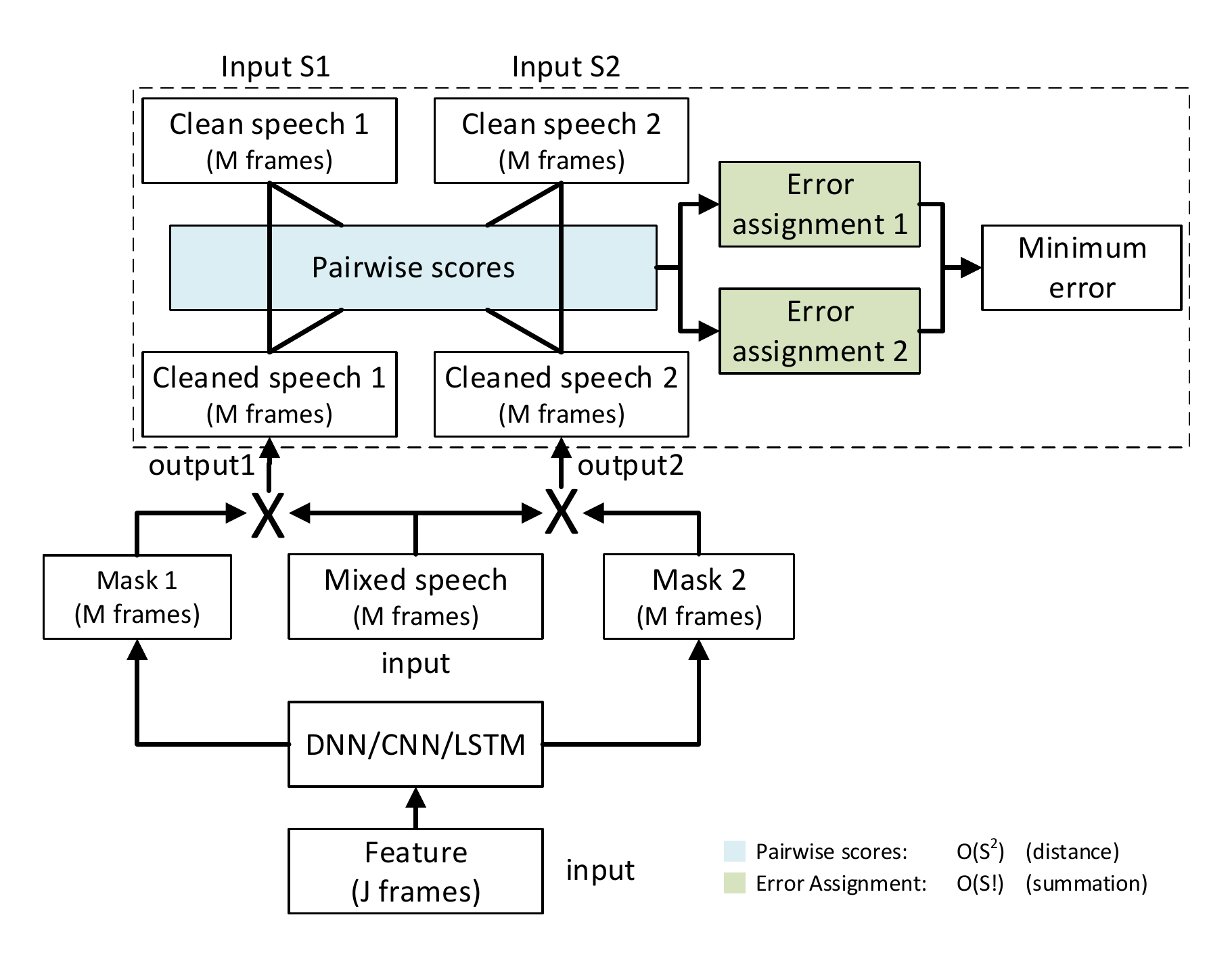}
		\caption{The two-talker speech separation model with permutation invariant training.}\label{fig:model}
	\end{figure}

	Our solution to the label permutation problem is illustrated in Fig.~\ref{fig:model} and is referred to as Permutation Invariant Training\;(PIT) \cite{yu_permutation_2017}. 

	In the model depicted in Fig.~\ref{fig:model} (and unlike the conventional model in Fig.\;\ref{fig:conventional}) the reference signals are given as \emph{a set} instead of an ordered list. In other words, the same training result is obtained, no matter in which order these references are listed. This behavior is achieved with PIT highlighted inside the dashed rectangle in Fig.~\ref{fig:model}. 
	Specifically, following the notation from Sec.\;\ref{sec:problem}, we associate the reference signals for speaker one and two, i.e. $\mathbf{a}_{1,i}$ and $\mathbf{a}_{2,i}$, to the output masks $\hat{\mathbf{m}}_{1,i}$ and $\hat{\mathbf{m}}_{2,i}$, by computing the (total of $S^2$) pairwise MSE's between each reference signal $\mathbf{a}_{s,i}$ and each estimated source $\hat {\mathbf{a}}_{s,i}$. We then determine the (total of $S!$) possible permutations between the references and the estimated sources, and compute the \emph{per-permutation-loss} for each permutation. 
	That is, for the two-speaker case in Fig.~\ref{fig:model} we compute the \emph{per-permutation-loss} for the two candidate output vectors $\hat{\mathbf{u}}_{i} = \left[ \hat{\mathbf{m}}_{1,i}^T \; \hat{\mathbf{m}}_{2,i}^T \right]^T$ and $\hat{\mathbf{u}}_{i} = \left[ \hat{\mathbf{m}}_{2,i}^T \; \hat{\mathbf{m}}_{1,i}^T \right]^T$.   
	The permutation with the lowest MSE is chosen and the model is optimized to reduce this least MSE. In other words, we simultaneously conduct label assignment and error evaluation. Similarly to prior arts, we can use $J$, and $M$ successive input, and output frames, respectively, (i.e., a {\em meta-frame}) to exploit the contextual information. 
    Note that only $S^2$ pairwise MSE's are required (and not $S!$) to compute the \emph{per-permutation-loss} for all $S!$ possible permutations. Since $S!$ grows much faster than $S^2$, with respect to $S$, and the computational complexity of the pairwise MSE is much larger than the \emph{per-permutation-loss} (sum of pairwise MSE's), PIT can be used with a large number of speakers, i.e. $S\gg2$.
	
	During inference, the only information available is the mixed speech, but speech separation can be directly carried out for each input meta-frame, for which an output meta-frame with $M$ frames of speech is estimated. Due to the PIT training criterion, the permutation will stay the same for frames inside the same output meta-frame, but may change across output meta-frames. 
	In the simplest setup, we can just assume that permutations do not change across output meta-frames, when reconstructing the target speakers. However, this usually leads to unsatisfactory results as reported in \cite{yu_permutation_2017}. To achieve better performance, speaker tracing algorithms, that identify the permutations of output meta-frames with respect to the speakers, need to be developed and integrated into the PIT framework or applied on top of the output of the network.

	\section{Utterance-level PIT}\label{sec:integrated}
	Several ways exist for identifying the permutation of the output meta-frames, i.e. solving the tracing problem. 
	For example, in CASA a related problem referred to as the Sequential Organization Problem has been addressed using a model-based sequential grouping algorithm \cite{shao_model-based_2006}. Although moderately successful for co-channel speech separation, where prior knowledge about the speakers is available, this method is not easily extended to the speaker independent case with multiple speakers. Furthermore, it is not easily integrated into a deep learning framework.      
	
	A more straight-forward approach might be to determine a change in permutation by comparing MSEs for different permutations of output masks measured on the overlapping frames of adjacent output meta-frames. However, this approach has two major problems. First, it requires a separate tracing step, which may complicate the model. Second, since the permutation of later frames depends on that of earlier frames, one incorrect assignment at an earlier frame would completely switch the permutation for all frames after it, even if the assignment decisions for the remaining frames are all correct.

	In this work we propose utterance-level Permutation Invariant Training\;(uPIT), a simpler yet more effective approach to solve the tracing problem and the label permutation problem than original PIT. Specifically, we extend the frame-level PIT technique with the following utterance-level cost function: 
	\begin{equation}
	J_{\phi^\ast} = \frac{1}{B}\sum_{s=1}^S \|\hat{\mathbf{M}}_s  \circ \mathbf{R} - \mathbf{A}_{\phi^\ast(s)}  \circ \cos(\mathbf{\theta}_y - \mathbf{\theta}_{\phi^\ast(s)})\|_F^2,	
	\label{eqPITutt1}
	\end{equation}	
	where $\phi^\ast$ is the permutation that minimizes the utterance-level separation error defined as
	\begin{equation}
	\phi^\ast = \underset{\phi\in \mathcal{P}}{\text{argmin}} \sum_{s=1}^S  \| \hat{\mathbf{M}}_s  \circ \mathbf{R} -  \mathbf{A}_{\phi(s)}\circ \cos(\mathbf{\theta}_y - \mathbf{\theta}_{\phi(s)}) \|_F^2,
	\label{eqPITutt}
	\end{equation}
	and $\mathcal{P}$ is the symmetric group of degree $S$, i.e. the set of all $S!$ permutations.

	In original PIT, the optimal permutation (in MSE sense) is computed and applied \emph{for each} output meta-frame. This implies that consecutive meta-frames might be associated with different permutations, and although PIT solves the label permutation problem, it does not solve the speaker tracing problem.  
	With uPIT, however, the permutation corresponding to the minimum utterance-level separation error is used \emph{for all} frames in the utterance. 
	In other words, the pair-wise scores in Fig.~\ref{fig:model} are computed for the whole utterance assuming all output frames follow the same permutation.
	Using the same permutation \emph{for all} frames in the utterance might imply that a non-MSE-optimal permutation is used for individual frames within the utterance. However, the intuition behind uPIT is that since the permutation resulting in the minimum utterance-level separation error is used, the number of non-optimal permutations is small and the model sees enough correctly permuted frames to learn an efficient separation model.
	For example, the output vector $\hat{\mathbf{u}}_{i}$ of a perfectly trained two-talker speech separation model, given an input utterance, should ideally be $\hat{\mathbf{u}}_{i} = \left[ \hat{\mathbf{m}}_{1,i}^T \; \hat{\mathbf{m}}_{2,i}^T \right]^T$, or $\hat{\mathbf{u}}_{i} = \left[ \hat{\mathbf{m}}_{2,i}^T \; \hat{\mathbf{m}}_{1,i}^T \right]^T \forall \; i=1, \dots , T$, i.e. the output masks should follow the same permutation for all $T$ frames in the utterance. 
	Fortunately, using Eq.\;\eqref{eqPITutt1} as a training criterion, for deep learning based speech separation models, this seems to be the case in practice (See Sec.\;\ref{sec:exp} for examples). 

	Since utterances have variable length, and effective separation presumably requires exploitation of long-range signal dependencies, models such as DNNs and CNNs are no longer good fits. Instead, we use deep LSTM RNNs and bi-directional LSTM (BLSTM) RNNs together with uPIT to learn the masks.
	Different from PIT, in which the input layer and each output layer has $N \times T$  and  $N \times M$ units, respectively, in uPIT, both input and output layers have $N$ units (adding contextual frames in the input does not help for LSTMs). With deep LSTMs, the utterance is evaluated frame-by-frame exploiting the whole past history information at each layer. When BLSTMs are used, the information from the past and future (i.e., across the whole utterance) is stacked at each layer and used as the input to the subsequent layer. 
	With uPIT, during inference we don't need to compute pairwise MSEs and errors of each possible permutation and no additional speaker tracing step is needed. We simply assume a constant permutation and treat the same output mask to be from the same speaker for all frames. This makes uPIT a simple and attractive solution.

	\section{Experimental results}
	\label{sec:exp}
	
	We evaluated uPIT on various setups and all models were implemented using the Microsoft Cognitive Toolkit (CNTK) \cite{yu_computational_2015,agarwal_introduction_2014}\footnote{Available at: https://www.cntk.ai/}. The models were evaluated on their potential to improve the Signal-to-Distortion Ratio\;(SDR) \cite{vincent_performance_2006} and the Perceptual Evaluation of Speech Quality\;(PESQ) \cite{rix_perceptual_2001} score, both of which are metrics widely used to evaluate speech enhancement performance for multi-talker speech separation tasks.
	
	\subsection{Datasets}
	\label{subsec:datasets}
	
	We evaluated uPIT on the WSJ0-2mix, WSj0-3mix\footnote{Available at: http://www.merl.com/demos/deep-clustering} and Danish-2mix datasets using 129-dimensional STFT magnitude spectra computed with a sampling frequency of 8 kHz, a frame size of 32 ms and a 16 ms frame shift.
	
	The WSJ0-2mix dataset was introduced in \cite{hershey_deep_2016} and was derived from the WSJ0 corpus \cite{garofolo_csr-i_1993}. The 30h training set and the 10h validation set contain two-speaker mixtures generated by randomly selecting from 49 male and 51 female speakers and utterances from the WSJ0 training set si\_tr\_s, and mixing them at various Signal-to-Noise Ratios\;(SNRs) uniformly chosen between 0\,dB and 5\,dB. The 5h test set was similarly generated using utterances from 16 speakers from the WSJ0 validation set si\_dt\_05 and evaluation set si\_et\_05. The WSJ0-3mix dataset was generated using a similar approach but contains mixtures of speech from three talkers. 
	
	The Danish-2mix dataset is based on a corpus\footnote{Available at: {\url{http://www.nb.no/sbfil/dok/nst_taledat_dk.pdf}} } with  approximately 560 speakers each speaking 312 utterances with average utterance duration of approximately 5 sec. The dataset was constructed by randomly selecting a set of 45 male and 45 female speakers from the corpus, and then allocating 232 and 40 utterances from each speaker to generate mixed speech in the training, and validation set, respectively. A number of 40 utterances from each of another 45 male and 45 female speakers were randomly selected to construct the open-condition\;(OC) (unseen speaker) test set. Speech mixtures were constructed similarly to the WSJ0-2mix with SNRs selected uniformly between 0\,dB and 5\,dB. Similarly to the WSJ0-2mix dataset we constructed 20k and 5k mixtures in total in the training and validation set, respectively, and 3k mixtures for the OC test set. 
	
	In our study, the validation set is used to find initial hyperparameters and to evaluate closed-condition\;(CC) (seen speaker) performance, similarly to \cite{hershey_deep_2016,isik_single-channel_2016,yu_permutation_2017}.
	
	\subsection{Permutation Invariant Training}

	We first evaluated the original frame-level PIT on the two-talker separation dataset WSJ0-2mix, and differently from \cite{yu_permutation_2017}, we fixed the input dimension to 51 frames, to isolate the effect of a varying output dimension.
	In PIT, the input window and output window sizes are fixed. For this reason, we can use DNNs and CNNs. The DNN model has three hidden layers each with 1024 ReLU units. In (inChannel, outChannel)-(strideW, strideH) format, the CNN model has one $(1,64)-(2,2)$, four $(64,64)-(1,1)$, one $(64,128)-(2,2)$, two $(128,128)-(1,1)$, one $(128,256)-(2,2)$, and two $(256,256)-(1,1)$ convolution layers with $3 \times 3$ kernels, a $7\times17$ average pooling layer and a 1024-unit ReLU layer. The input to the models is the stack (over multiple frames) of the 129-dimensional STFT spectral magnitude of the speech mixture. 
	The output layer $\hat{\mathbf{u}}_{i}$ is divided into $S$ output masks/streams for $S$-talker mixed speech as $\hat{\mathbf{u}}_{i} = \left[ \hat{\mathbf{m}}_{1,i} \;;\; \hat{\mathbf{m}}_{2,i} \;;\; \dots \;;\; \hat{\mathbf{m}}_{S,i} \right]^T$. Each output mask vector $\hat{\mathbf{m}}_{s,i}$ has a dimension of $129 \times M$, where $M$ is the number of frames in the output meta-frame.

	\begin{figure}[ht] 
	\centering
	\centerline{\includegraphics[trim={8mm 0mm 8mm 2mm},width=1\linewidth]{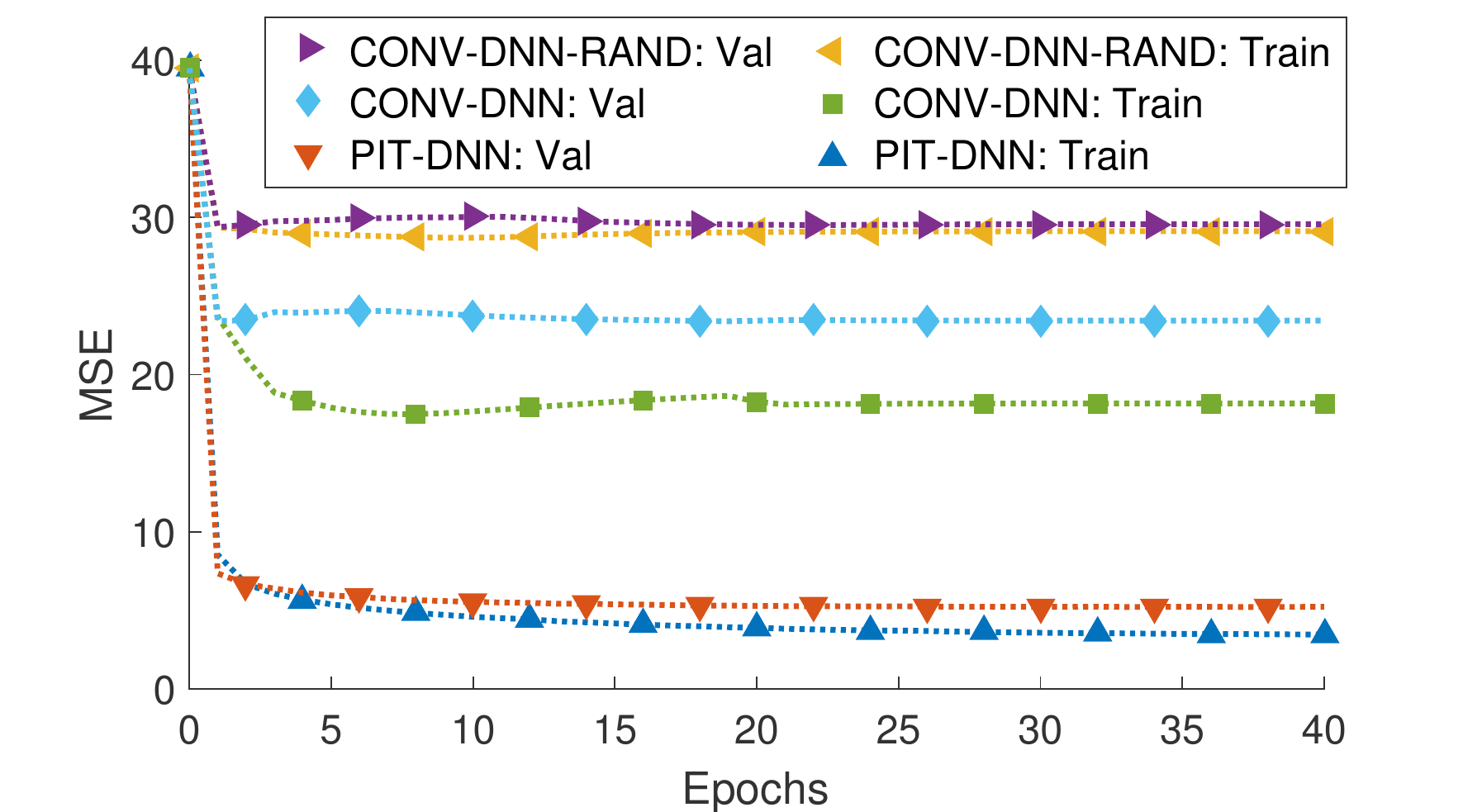}}
		\caption{MSE over epochs on the WSJ0-2mix training and validation sets with conventional training and PIT.}\label{fig:mse}
	\end{figure}

	In Fig.~\ref{fig:mse} we present the DNN training progress as measured by the MSE on the training and validation set with conventional training (CONV-DNN) and PIT on the WSJ0-2mix datasets described in subsection \ref{subsec:datasets}. We also included the training progress for another conventionally trained model but with a slightly modified version of the WSJ0-2mix dataset, where speaker labels have been randomized (CONV-DNN-RAND). 
	
	The WSJ0-2mix dataset, used in \cite{hershey_deep_2016}, was designed such that speaker one was always assigned the most energy, and consequently speaker two the lowest, when scaling to a given SNR. 
	Previous work \cite{weng_deep_2015} has shown that such speaker energy patterns are an effective discriminative feature, which is clearly seen in Fig.~\ref{fig:mse}, where the CONV-DNN model achieves considerably lower training and validation MSE than the CONV-DNN-RAND model, which hardly decreases in either training or validation MSE due to the label permutation problem \cite{weng_deep_2015,hershey_deep_2016}. 
	In contrast, training converges quickly to a very low MSE when PIT is used.
	\begin{table}[t]
		\caption{SDR improvements (dB) for different separation methods on the WSJ0-2mix dataset using PIT.}
		\label{tab:WSJ0-2mix-PIT}
		\centering
		\begin{tabular}{l|c|cc|cc}
			\toprule
			Method & Input\textbackslash Output & \multicolumn{2}{c|} {Opt. Assign.} & \multicolumn{2}{c} {Def. Assign.}\\ 
			&  window                    		& CC & OC		&  CC 	& OC	\\
			\midrule
			PIT-DNN    & 51\textbackslash51    		& 6.8 & 6.7  & \bf{5.2} & \bf{5.2}     \\ 
			PIT-DNN    & 51\textbackslash5    		& \bf{10.3} & \bf{10.2}  & -0.8 & -0.8     \\ 
			\midrule
			PIT-CNN    & 51\textbackslash51     	& 9.6 & 9.6  & \bf{7.6} & \bf{7.5}     \\
			PIT-CNN    & 51\textbackslash5      	& \bf{10.9} & \bf{11.0}  & -1.0 & -0.9     \\
			\midrule
			IRM 	& - & 12.4 & 12.7 & 12.4 & 12.7 \\
			IPSM 	& - & 14.9 & 15.1 & 14.9 & 15.1 \\
			\bottomrule
		\end{tabular}
	\end{table}
	
	In Table~\ref{tab:WSJ0-2mix-PIT} we summarize the SDR improvement in dB from different frame-level PIT separation configurations for two-talker mixed speech in closed condition\;(CC) and open condition\;(OC). In these experiments each frame was reconstructed by averaging over all output meta-frames that contain the same frame. In the default assignment (def. assign.) setup, a constant output mask permutation is assumed across frames (which is an invalid assumption in general). 
	This is the maximum achievable SDR improvement using PIT without the utterance-level training criterion and without an additional tracing step. 
	In the optimal assignment (opt. assign.) setup, the output-mask permutation for each output meta-frame is determined based on the true target, i.e. oracle information. This reflects the separation performance within each segment (meta-frame) and is the improvement achievable when the speakers are correctly separated. The gap between these two values indicates the possible contribution from speaker tracing. As a reference, we also provided the IRM and IPSM results.

	From the table we can make several observations. First, PIT can already achieve 7.5 dB SDR improvement (def. assign.), even though the model is very simple. Second, as we reduce the output window size, we can improve the separation performance within each window and achieve better SDR improvement, if speakers are correctly traced (opt. assign.). However, when output window size is reduced, the output mask permutation changes more frequently as indicated by the poor default assignment performance. Speaker tracing thus becomes more important given the larger gap between the optimal assignment and default assignment. Third, PIT generalizes well to unseen speakers, since the performances on the open and closed conditions are very close. Fourth, powerful models such as CNNs consistently outperform DNNs, but the gain diminishes when the output window size is small.  
	
	\subsection{Utterance-level Permutation Invariant Training}
	\begin{table}[t]
		\caption{SDR improvements (dB) for different separation methods on the WSJ0-2mix dataset using uPIT.}
		\label{tab:WSJ0-2mix-uPIT}
		\centering
		\begin{tabular}{l|c|c|cc|cc}
			\toprule
			Method & Mask &  Activation & \multicolumn{2}{c|} {Opt. Assign.} & \multicolumn{2}{c} {Def. Assign.}\\ 
			&  Type                    		& Function & CC		&  OC 	& CC & OC	\\
			\midrule
			uPIT-BLSTM    & AM & softmax   		& \bf{10.4} &     \bf{10.3}	& 	\bf{9.0} 	& 	\bf{8.7}  \\  
			uPIT-BLSTM    & AM & sigmoid      	& 8.3 		&     8.3  	&   7.1 		&   7.2     \\ 
			uPIT-BLSTM    & AM & ReLU     		& 9.9 		&     9.9      &   8.7 		&   8.6     \\ 
			uPIT-BLSTM    & AM & Tanh      	    & 8.5  	    &     8.6      &   7.5 		&   7.5     \\ 
			\midrule
			uPIT-BLSTM    & PSM & softmax   		& 10.3 		&     10.2  	& 	9.1 		& 	9.0     \\
			uPIT-BLSTM    & PSM & sigmoid      	& 10.5 		&     10.4  	& 	9.2 		& 	9.1     \\
			uPIT-BLSTM    & PSM & ReLU     		& \bf{10.9}	&     \bf{10.8}	& 	\bf{9.4} 	& 	\bf{9.4}     \\
			uPIT-BLSTM    & PSM & Tanh      	    & 10.4 		&     10.3  	& 	9.0 		& 	8.9     \\ 
			\midrule
			uPIT-BLSTM    & NPSM & softmax   		& 8.7 		&     8.6  	& 	7.5 		& 	7.3     \\
			uPIT-BLSTM    & NPSM & sigmoid      	& \bf{10.6}	& \bf{10.6}	& 	\bf{9.4} 	& 	\bf{9.3}     \\
			uPIT-BLSTM    & NPSM & ReLU     		& 8.8 		&     8.8  	& 	7.6 		& 	7.6     \\
			uPIT-BLSTM    & NPSM & Tanh      	    & 10.1 		&     10.0  	& 	8.9 		& 	8.8     \\
			\midrule
			uPIT-LSTM    & PSM & ReLU   			& \bf{9.8} & \bf{9.8} & 7.0 & \bf{7.0} \\ 
			uPIT-LSTM    & PSM & sigmoid      		& \bf{9.8} & 9.6	& \bf{7.1} & 6.9 \\ 
			uPIT-LSTM    & NPSM & ReLU     		& \bf{9.8} & \bf{9.8} & \bf{7.1} & \bf{7.0} \\ 
			uPIT-LSTM    & NPSM & sigmoid      	& 9.2 		& 9.2 	&  6.8 		& 6.8 \\ 
			\midrule
			PIT-BLSTM    & PSM & ReLU   			& \bf{11.7} & \bf{11.7} & \bf{-1.7} & -1.9   \\ 
			PIT-BLSTM    & PSM & sigmoid      	& \bf{11.7} & \bf{11.7} & \bf{-1.7} & \bf{-1.7} \\ 
			PIT-BLSTM    & NPSM & ReLU     		& \bf{11.7} & \bf{11.7} & \bf{-1.7} & -1.8 \\ 
			PIT-BLSTM    & NPSM & sigmoid      	& 11.6 & 11.6 & -1.6 & \bf{-1.7} \\ 
			\midrule
			IRM  & - & -  & 12.4 & 12.7 & 12.4 & 12.7 \\
			IPSM & - & -  & 14.9 & 15.1 & 14.9 & 15.1 \\
			\bottomrule
		\end{tabular}
		\vspace{-0.25mm}
	\end{table}
	
	As indicated by Table~\ref{tab:WSJ0-2mix-PIT}, an accurate output mask permutation is critical to further improve the separation quality. In this subsection we evaluate the uPIT technique as discussed in Sec.~\ref{sec:integrated} and the results are summarized in Table~\ref{tab:WSJ0-2mix-uPIT}. 
	
	Due to the formulation of the uPIT cost function in Eq.\;\eqref{eqPITutt1} and Eq.\;\eqref{eqPITutt}, and to utilize long-range context, RNNs are the natural choice, and in this set of experiments, we used LSTM RNNs. All the uni-directional LSTMs (uPIT-LSTM) evaluated have 3 LSTM layers each with 1792 units and all the bi-directional LSTMs (uPIT-BLSTM) have 3 BLSTM layers each with 896 units, so that both models have similar number of parameters. 
	
	All models contain random dropouts when fed from a lower layer to a higher layer and were trained with a dropout rate of 0.5. Note that, since we used Nvidia's cuDNN implementation of LSTMs, to speed up training, we were unable to apply dropout across time steps, which was adopted by the best DPCL model \cite{isik_single-channel_2016} and is known to be more effective, both theoretically and empirically, than the simple dropout strategy used in this work \cite{gal_theoretically_2015}. 
	
	In all the experiments reported in Table~\ref{tab:WSJ0-2mix-uPIT} the maximum epoch is set to 200 although we noticed that further performance improvement is possible with additional training epochs. Note that the epoch size of 200 seems to be significantly larger than that in PIT as indicated in Fig.~\ref{fig:mse}. This is likely because in PIT each frame is used by $T$ ($T=51$) training samples (input meta-frames) while in uPIT each frame is used just once in each epoch. 
	
	The learning rates were set to $2 \times 10^{-5}$ per sample initially and scaled down by $0.7$ when the training objective function value increases on the training set. The training was terminated when the learning rate got below $10^{-10}$. Each minibatch contains 8 randomly selected utterances.

	As a related baseline, we also include PIT-BLSTM results in Table~\ref{tab:WSJ0-2mix-uPIT}. These models were also trained using LSTMs with whole utterances instead of meta-frames. The only difference between these models and uPIT models is that uPIT models use the utterance-level training criterion defined in Eqs.\,\eqref{eqPITutt1} and \eqref{eqPITutt}, instead of the meta-frame based criterion used by PIT.
	%
	\subsubsection{uPIT Training Progress}
	\begin{figure}[ht] 
	\centering
	\centerline{\includegraphics[trim={8mm 0mm 8mm 2mm},width=1.0\linewidth]{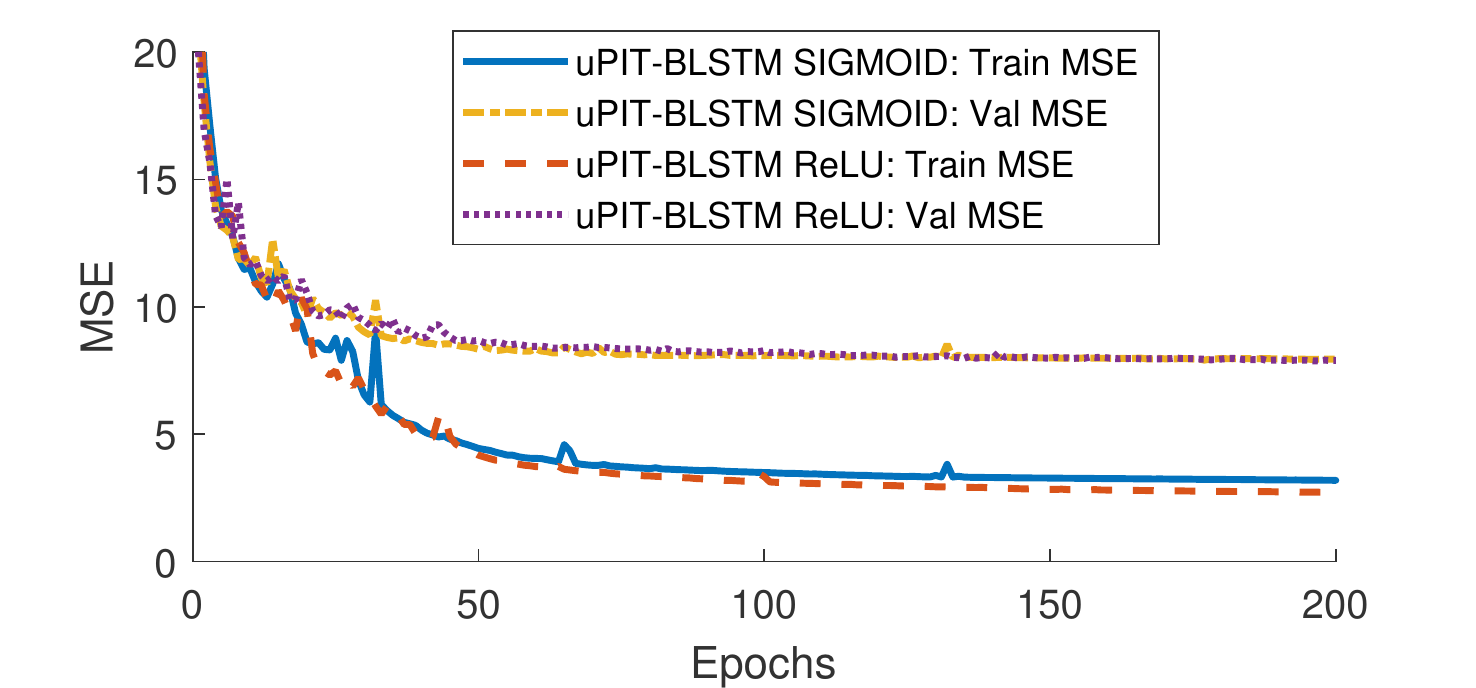}}
		\caption{MSE over epochs on the WSJ0-2mix PSM training and validation sets wit uPIT.}\label{fig:mse-uPIT}
	\end{figure}
	
	In Fig.~\ref{fig:mse-uPIT} we present a representative example of the BLSTM training progress, as measured by the MSE of the two-talker mixed speech training and validation set, using Eq.\;\eqref{eqPITutt1}. We see that the training and validation MSE's are both steadily decreasing as function of epochs, hence uPIT, similarly to PIT, effectively solves the label permutation problem.
	
	\subsubsection{uPIT Performance for Different Setups}
	From Table~\ref{tab:WSJ0-2mix-uPIT}, we can notice several things. First, with uPIT, we can significantly improve the SDR with default assignment over original PIT. In fact, a 9.4\,dB SDR improvement on both CC and OC sets can be achieved by simply assuming a constant output mask permutation (def. assign.), which compares favorably to 7.6\,dB (CC) and 7.5\,dB (OC) achieved with deep CNNs combined with PIT. 
	We want to emphasize that this is achieved through Eqs.\;\eqref{eqPITutt1} and \eqref{eqPITutt}, and not by using BLSTMs because the corresponding PIT-BLSTM default assignment results are so much worse, even though the optimal assignment results are the best among all models. 
	The latter may be explained from the PIT objective function that attempts to obtain a constant output mask permutation at the meta-frame-level, which for small meta-frames is assumed easier compared to the uPIT objective function, that attempts to obtain a constant output mask permutation throughout the whole utterance. 
	Second, we can achieve better SDR improvement over the AM using PSM and NPSM training criteria. 
	This indicates that including phase information does improve performance, even-though it was used implicitly via the cosine term in Eq.\;\eqref{eqPITutt1}. 
	Third, with uPIT the gap between optimal assignment and default assignment is always less than 1.5\,dB across different setups, hence additional improvements from speaker tracing algorithms is limited to 1.5\,dB. 	
	\subsubsection{Two-stage Models and Reduced Dropout Rate}
	It is well known that cascading DNNs can improve performance for certain deep learning based applications \cite{zhang_deep_2016,nie_deep_2014,wang_recurrent_2017,isik_single-channel_2016}.
	In Table~\ref{tab:WSJ0-2mix-Further} we show that a similar principle of cascading two BLSTM models into a two-stage model (-ST models in Table~\ref{tab:WSJ0-2mix-Further}) can lead to improved performance over the models presented in Table~\ref{tab:WSJ0-2mix-uPIT}. 
	In Table~\ref{tab:WSJ0-2mix-Further} we also show that improved performance, with respect to the same models, can be achieved with additional training epochs combined with a reduced dropout rate (-RD models in Table~\ref{tab:WSJ0-2mix-Further}).
	Specifically, if we continue the training of the two best performing models from Table~\ref{tab:WSJ0-2mix-uPIT} (i.e. uPIT-BLSTM-PSM-ReLU and uPIT-BLSTM-NPSM-Sigmoid) with 200 additional training epochs at a reduced dropout rate of 0.3, we see an improvement of $0.1$\;dB.
	Even larger improvements can be achieved with the two-stage approach, where an estimated mask is computed as the average mask from two BLSTM models as 
	\begin{equation}
	\begin{split}
	\hat{\mathbf{M}}_s &= \frac {\hat{\mathbf{M}}_s^{(1)} + \hat{\mathbf{M}}_s^{(2)}} {2}.
	\end{split}
	\end{equation}
	The mask $\hat{\mathbf{M}}_s^{(1)}$ is from an -RD model that serves as a first-stage model, and $\hat{\mathbf{M}}_s^{(2)}$ is the output mask from a second-stage model. The second-stage model is trained using the original input features as well as the mask $\hat{\mathbf{M}}_s^{(1)}$ from the first-stage model.
	The intuition behind this architecture is that the second-stage model will learn to correct the errors made by the first-stage model.   
	\begin{table}[t]
		\caption{Further improvement on the WSJ0-2mix dataset with additional training epochs with reduced dropout (-RD) or stacked models (-ST)}
		\label{tab:WSJ0-2mix-Further}
		\centering
		\begin{tabular}{l|c|c|cc|cc}
			\toprule
			Method & Mask &  Activation   & \multicolumn{2}{c|} {Opt. Assign.} & \multicolumn{2}{c} {Def. Assign.}\\ 
			&  Type                & Function     		& CC & OC		&  CC 	& OC	\\
			\midrule
			uPIT-BLSTM-RD  & PSM 	& ReLU  	   	& 11.0 	&     11.0 	& 9.5 	& 9.5     \\ 
			uPIT-BLSTM-ST  & PSM 	& ReLU   	    & \bf{11.7} & \bf{11.7} & {10.0} & \bf{10.0}     \\ 
			uPIT-BLSTM-RD  & NPSM 	& Sigmoid  	   	& 10.7 	&     10.7  	& 9.5 	& 9.4     \\ 
			uPIT-BLSTM-ST  & NPSM 	& Sigmoid       & {11.5} 	& {11.5}  	& \bf{10.1} & \bf{10.0}     \\  
			\midrule
			IRM 		   & -  	& -  			& 12.4 & 12.7 & 12.4 & 12.7 \\
			IPSM 		   & - 		& -				& 14.9 & 15.1 & 14.9 & 15.1 \\
			\bottomrule
		\end{tabular}
		\vspace{-0.25mm}
	\end{table}
	Table~\ref{tab:WSJ0-2mix-Further} shows that the two-stage models (-ST models) always outperform the single-stage models (-RD models) and overall, a 10\,dB SDR improvement can be achieved on this task using a two-stage approach.

	\subsubsection{Opposite Gender vs. Same Gender.}
	\begin{table}[t]
		\caption{SDR (dB) improvements on test sets of WSJ0-2mix divided into same and opposite gender mixtures}
		\label{tab:WSJ0-2mix-gender}
		\centering
		\begin{tabular}{l|c|cc|cc}
			\toprule
			Method & Config & \multicolumn{2}{c|} {CC} & \multicolumn{2}{c} {OC}\\ 
			&                      		& Same & Opp.		&  Same 	& Opp.	\\
			\midrule
			uPIT-BLSTM-RD   & PSM-ReLU    		& 7.5 	& 11.5  	& 7.1 	& 11.6     \\ 
			uPIT-BLSTM-ST	 & PSM-ReLU   		& 7.8 	& \bf{12.1} & \bf{7.5} & \bf{12.2}    \\ 
			uPIT-BLSTM-RD   & NPSM-Sigmoid    	& 7.5 	& 11.5  	& 7.0 	& 11.5     \\ 
			uPIT-BLSTM-ST   & NPSM-Sigmoid   	& \bf{8.0} & \bf{12.1}  	& \bf{7.5} 	& 12.1  \\ 
			\midrule
			IRM  	& -							& 12.2  	& 12.7	& 12.4 	& 12.9   \\ 
			IPSM 	& - 						& 14.6 		& 15.1 	& 14.9 	& 15.3 \\
			\bottomrule
		\end{tabular}
	\end{table}
	
	Table~\ref{tab:WSJ0-2mix-gender} reports SDR (dB) improvements on test sets of WSJ0-2mix divided into opposite-gender\;(Opp.) and same-gender\;(Same). From this table we can clearly see that our approach achieves much better SDR improvements on the opposite-gender mixed speech than the same-gender mixed speech, although the gender information is not explicitly used in our model and training procedure. In fact, for the opposite-gender condition, the SDR improvement is already very close to the IRM result. These results agree with breakdowns from other works \cite{hershey_deep_2016,isik_single-channel_2016} and generally indicate that same-gender mixed speech separation is a harder task.

	\subsubsection{Multi-Language Models}
	To further understand the properties of uPIT, we evaluated the uPIT-BLSTM-PSM-ReLU model trained on WSJ0-2mix (English) on the Danish-2mix test set. The results of this is reported in Table~\ref{tab:twolanguage}. 
	\begin{table}[t]
		\caption{SDR (dB) and PESQ improvements on WSJ0-2mix and Danish-2mix with uPIT-BLSTM-PSM-ReLU trained on WSJ0-2mix and a combination of two languages.}
		\label{tab:twolanguage}
		\centering
		\begin{tabular}{l|c|c|c|cc}
			\toprule
			Trained on & \multicolumn{2}{c|} {WSJ0-2mix} & \multicolumn{2}{c} {Danish-2mix}\\ 
			& SDR & PESQ		&  SDR & PESQ	\\
			\midrule
			WSJ0-2mix    	& 9.4 	&     0.62  	&     8.1 	&     0.40     \\ 
			+Danish-2mix    & 8.8 	&     0.58  	&     10.6	&     0.51     \\ 
			\midrule
			IRM	   		    & 12.7  & 2.11      	& 15.2  & 1.90  \\ 
			IPSM 			& 15.1 	& 2.10 		& 17.7 	& 1.90 \\
			\bottomrule
		\end{tabular}
		\vspace{-0.25mm}
	\end{table}
	An interesting observation, is that although the system has never seen Danish speech, it performs remarkably well in terms of SDR, when compared to the IRM (oracle) values. These results indicate, that the separation ability learned with uPIT generalizes well, not only across speakers, but also across languages. 
	In terms of PESQ, we see a somewhat larger performance gap with respect to the IRM. This might be explained by the fact that SDR is a waveform matching criteria and does not necessarily reflect perceived quality as well as PESQ. Furthermore, we note that the PESQ improvements are similar to what have been reported for DNN based speech enhancement systems \cite{kolbaek_speech_2017}.
	
	We also trained a model with the combination of English and Danish datasets and evaluated the models on both languages. The results of these experiments are summarized in Table~\ref{tab:twolanguage}. Table~\ref{tab:twolanguage}, indicate that by including Danish data, we can achieve better performance on the Danish dataset, at the cost of slightly worse performance on the English dataset. Note that while doubling the training set, we did not change the model size. Had we done this, performance would likely improve on both languages.

	\subsubsection{Summary of Multiple 2-Speaker Separation Techniques}   
		\begin{table}[t]
			\caption{SDR (dB) and PESQ improvements for different separation methods on the WSJ0-2mix dataset without additional tracing (i.e., def. assign.). $^\ddagger$ indicates curriculum training.}
			\label{tab:WSJ0-2mix-summary}
			\centering
			\begin{tabular}{l|c|cc|cc}
				\toprule
				Method & Config & \multicolumn{2}{c|} {PESQ Imp.} & \multicolumn{2}{c} {SDR Imp.}\\ 
				&                      		& CC & OC		&  CC 	& OC	\\
				\midrule
				Oracle NMF \cite{hershey_deep_2016}			& -	& -	& -		& 5.1  	& -     \\ 
				CASA \cite{hershey_deep_2016}   			& - & - & - 	& 2.9  	& 3.1   \\
				DPCL \cite{hershey_deep_2016} 				& - & - & -  	& 5.9  	& 5.8   \\
				DPCL+ \cite{chen_deep_2017}   				& - & -	& -		&  -  	& 9.1     \\ 
				DANet \cite{chen_deep_2017}   				& - & -	& -		&  -  	& 9.6     \\ 
				DANet$^\ddagger$ \cite{chen_deep_2017}  				& - & -	& -		&  -  	& 10.5     \\ 
				DPCL++ \cite{isik_single-channel_2016}   	& - & -	& -		&  -  	& 9.4     \\ 
				DPCL++$^\ddagger$ \cite{isik_single-channel_2016}   	& - & -	& -		&  -  	& \bf{10.8}     \\ 
				\midrule
				PIT-DNN     & 51\textbackslash51    				& 0.24 	&     0.23  	& 5.2 	& 5.2     \\ 
				PIT-CNN     & 51\textbackslash51   					& 0.52	&     0.50  	& 7.6 	& 7.6     \\ 
				uPIT-BLSTM   & PSM-ReLU    							& 0.66 	&     0.62  	& 9.4 	& 9.4     \\ 
				uPIT-BLSTM-ST& PSM-ReLU   							& {\bf{0.86}} & {\bf{0.82}} & {\bf{10.0}} & {10.0}  \\ 
				\midrule
				IRM  		& -     								& 2.15 	& 2.11 	& 12.4 	& 12.7     \\ 
				IPSM 		& - 									& 2.14 	& 2.10 	& 14.9 & 15.1 \\ 
				\bottomrule
			\end{tabular}
		\end{table}

	Table~\ref{tab:WSJ0-2mix-summary} summarizes SDR (dB) and PESQ improvements for different separation methods on the WSJ0-2mix dataset. From the table we can observe that the models trained with PIT already achieve similar or better SDR than the original DPCL \cite{hershey_deep_2016}, respectively, with DNNs and CNNs. 
	Using the uPIT training criteria, we improve on PIT and achieve comparable performance with DPCL+, DPCL++ and DANet models\footnote{\cite{isik_single-channel_2016,chen_deep_2017} did not use the SDR measure from \cite{vincent_performance_2006}. Instead a related variant called scale-invariant SNR was used.} 
	reported in \cite{isik_single-channel_2016,chen_deep_2017}, which used curriculum training \cite{bengio_curriculum_2009}, and recurrent dropout \cite{gal_theoretically_2015}. 
	Note that, both uPIT and PIT models are much simpler than DANet, DPCL, DPCL+, and DPCL++, because uPIT and PIT models do not require any clustering step during inference or estimation of attractor points, as required by DANet.

	\subsection{Three-Talker Speech Separation} 
	%
	\begin{figure}[ht] 
		\centering
		\centerline{\includegraphics[trim={8mm 0mm 8mm 2mm},clip,width=1.0\linewidth]{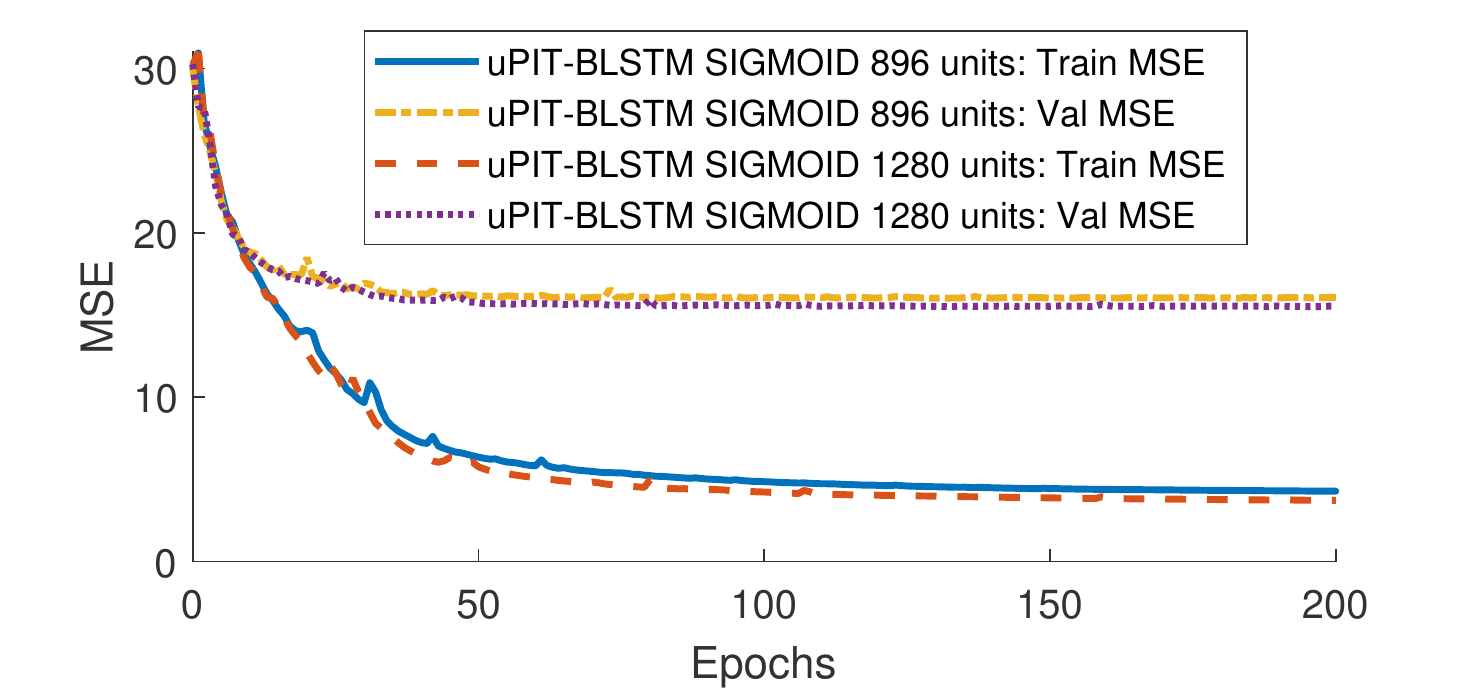}}
		\caption{MSE over epochs on the WSJ0-3mix NPSM training and validation sets wit uPIT.}\label{fig:mse-uPIT-3sprk}
	\end{figure}

	In Fig.~\ref{fig:mse-uPIT-3sprk} we present the uPIT training progress as measured by MSE on the three-talker mixed speech training and validation sets WSJ0-3mix. We observe that similar to the two-talker scenario in Fig.~\ref{fig:mse-uPIT}, a low training MSE is achieved, although the validation MSE is slightly higher. A better balance between the training and validation MSEs may be achieved by hyperparameter tuning. We also observe that increasing the model size decreases both training and validation MSE, which is expected due to the more variability in the dataset.   
	
	In Table~\ref{tab:WSJ0-3mix} we summarize the SDR improvement in dB from different uPIT separation configurations for three-talker mixed speech, in closed condition\;(CC) and open condition\;(OC).
	We observe that the basic uPIT-BLSTM model (896 units) compares favorably with DPCL++. 
	Furthermore, with additional units, further training and two-stage models (based on uPIT-BLSTM), uPIT achieves higher SDR than DPCL++ and similar SDR as DANet, without curriculum training, on this three-talker separation task.   

	\begin{table}[t]
		\caption{SDR improvements (dB) for different separation methods on the WSJ0-3mix dataset. $^\ddagger$ indicates curriculum training. }
		\label{tab:WSJ0-3mix}
		\centering
		\begin{tabular}{l|c|c|cc|cc}
			\toprule
			Method & Units/ &  Activation   & \multicolumn{2}{c|} {Opt. Assign.} & \multicolumn{2}{c} {Def. Assign.} \\ 
			&  layer & function     							& CC & OC		&  CC 	& OC	\\
			\midrule
			Oracle NMF \cite{hershey_deep_2016} & - & - & 4.5 & - & - & - \\
			DPCL++$^\ddagger$ \cite{isik_single-channel_2016} 		 & - & - & - & - & - & 7.1 \\ 
			DANet \cite{chen_single_2017}   				& - & -	& -	 & - & - & 7.7     \\ 
			DANet$^\ddagger$ \cite{chen_deep_2017}   				& - & -	& -	 & - & - & \bf{8.8}     \\ 
			\midrule
			uPIT-BLSTM		& 896 	& Sigmoid  	   	& 10.0 & 9.9 & 7.4 & 7.2 \\ 
			uPIT-BLSTM		& 1280 	& Sigmoid  	   	& 10.1 & 10.0 & 7.5 & 7.4 \\ 
			uPIT-BLSTM-RD	& 1280 	& Sigmoid  	   	& 10.2 & 10.1 & 7.6 & 7.4 \\ 
			uPIT-BLSTM-ST	& 1280 	& Sigmoid  	   	& \bf{10.7} & \bf{10.6} & \bf{7.9} & 7.7 \\ 
			\midrule
			IRM		& - 	& -  	   	& 12.6	 & 12.8 & 12.6 & 12.8 \\
			IPSM 	& - 	& - 		& 15.1 	& 15.3 	& 15.1 & 15.3 \\
			\bottomrule
		\end{tabular}
		\vspace{-0.25mm}
	\end{table}

	\subsection{Combined Two- and Three-Talker Speech Separation} 
	To illustrate the flexibility of uPIT, we summarize in Table~\ref{tab:WSJ0-2-3mix} the performance of the three-speaker uPIT-BLSTM, and uPIT-BLSTM-ST models (from Table~\ref{tab:WSJ0-3mix}), when they are trained and tested on both the WSJ0-2mix and WSJ0-3mix datasets, i.e. on both two- and three-speaker mixtures. 
	
	To be able to train the three-speaker models with the two-speaker WSJ0-2mix dataset, we extended WSJ0-2mix with a third "silent" channel. The silent channel consists of white Gaussian noise with an energy level 70 dB below the average energy level of the remaining two speakers in the mixture. 
	When we evaluated the model, we identified the two speaker-active output streams as the ones corresponding to the signals with the most energy. 
		
	We see from Table~\ref{tab:WSJ0-2-3mix} that uPIT-BLSTM achieves good, but slightly worse, performance compared to the corresponding two-speaker (Table~\ref{tab:WSJ0-2mix-summary}) and three-speaker (Table~\ref{tab:WSJ0-3mix}) models. 
	Surprisingly, the uPIT-BLSTM-ST model outperforms both the two-speaker (Table~\ref{tab:WSJ0-2mix-Further}) and three-speaker uPIT-BLSTM-ST (Table~\ref{tab:WSJ0-3mix}) models. These results indicate that a single model can handle a varying, and more importantly, unknown number of speakers, without compromising performance.    
	This is of great practical importance, since \emph{a priori} knowledge about the number of speakers is not needed at test time, as required by competing methods such as DPCL++ \cite{isik_single-channel_2016} and DANet \cite{chen_deep_2017,chen_single_2017}.  
	%
	
	\begin{table}[t]
		\caption{SDR improvements (dB) for three-speaker models trained on both the WSJ0-2mix and WSJ0-3mix PSM datasets. Both models have 1280 units per layer and ReLU outputs.}
		\label{tab:WSJ0-2-3mix}
		\centering
		\begin{tabular}{l|cc|cc}
			\toprule
			Method     & \multicolumn{2}{c|} {2 Spkr.} & \multicolumn{2}{c} {3 Spkr. } \\ 
			& \multicolumn{2}{c|} {Def. Assign.} & \multicolumn{2}{c} {Def. Assign.}	\\
			& CC & OC		                           &  CC 	& OC	\\
			\midrule
			uPIT-BLSTM		  	   	& 9.4 & 9.3 & 7.2 & 7.1 \\ 
			uPIT-BLSTM-ST	  	   	& \bf{10.2} & \bf{10.1} & \bf{8.0} 	& \bf{7.8} \\ 
			\midrule
			IRM  					& 12.4 		& 12.7 		& 12.6 & 12.8 \\
			IPSM    				& 14.9 		& 15.1 		& 15.1 & 15.3 \\
			\bottomrule
		\end{tabular}
		\vspace{-0.3mm}
	\end{table}
	
\begin{figure}[ht] 
		\centering
		\centerline{\includegraphics[trim={7mm 14mm 12mm 12mm},clip,width=0.95\linewidth]{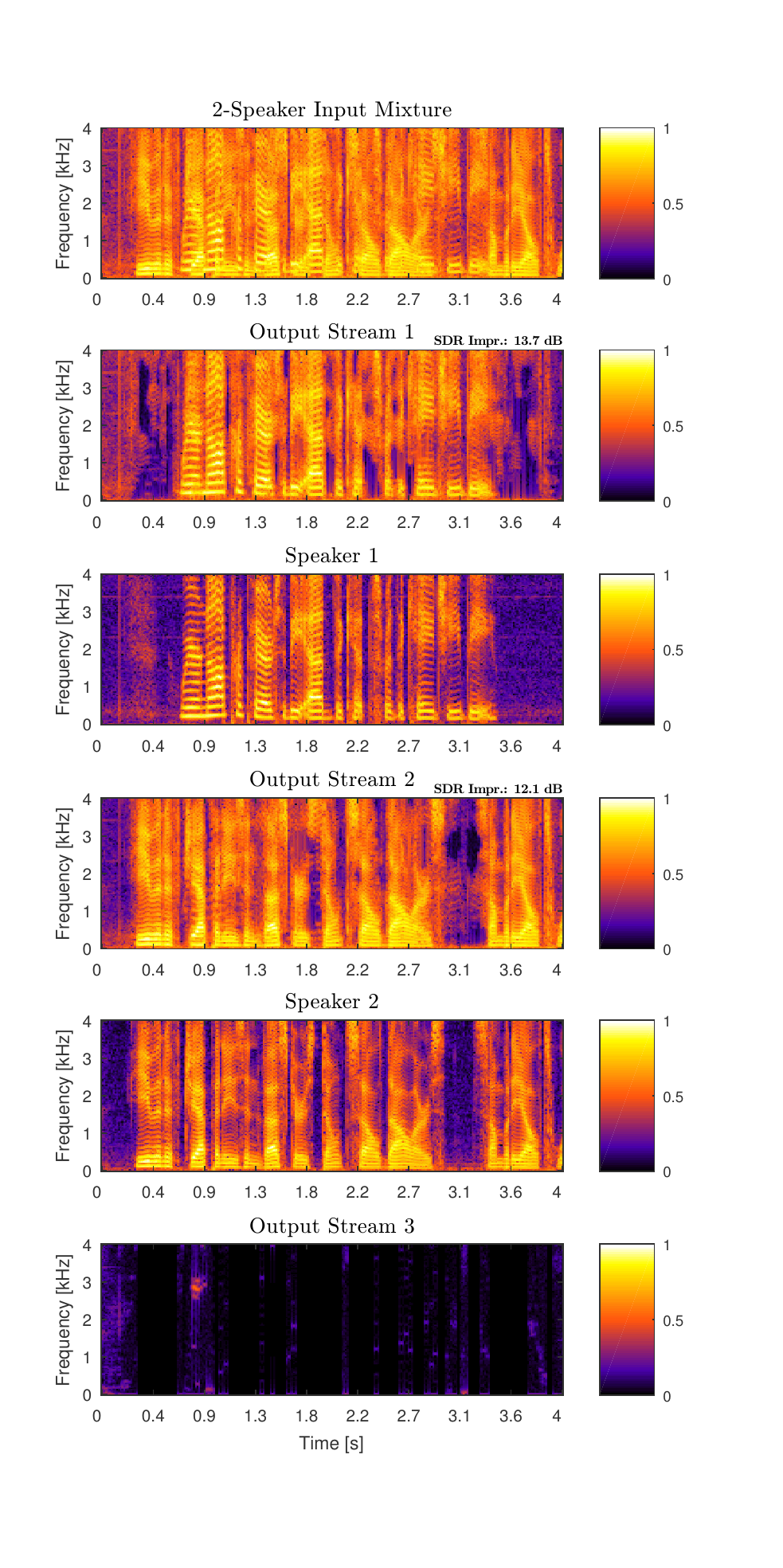}}
		\caption{Spectrograms showing how a three-speaker BLSTM model trained with uPIT can separate a two-speaker mixture while keeping a constant output-mask permutation. The energy in output stream three is $63$ dB lower than the energy in output stream one and two.}\label{fig:multSpkr}
	\end{figure}

	During evaluation of the 3000 mixtures in the WSJ0-2mix test set, output stream one and two were the output streams with the most energy, i.e. the speaker-active output streams, in 2999 cases. Furthermore, output stream one and two had, on average, an energy level approximately 33 dB higher than the silent channel, indicating that the models successfully keep a constant permutation of the output masks throughout the test utterance. 
	As an example, Fig.~\ref{fig:multSpkr} shows the spectrogram for a single two-speaker (male-vs-female) test case along with the spectrograms of the three output streams of the uPIT-BLSTM model, as well as the clean speech signals from each of the two speakers. 
	Clearly, output streams one and two contain the most energy and output stream three consists primarily of a low energy signal without any clear structure. 
	Furthermore, by comparing the spectrograms of the clean speech signals ("Speaker 1" and "Speaker 2" in Fig.~\ref{fig:multSpkr}) to the spectrogram of the corresponding output streams, it is observed that they share many similarities, which indicate that the model kept a constant output-mask permutation for the entire mixture and successfully separated the two speakers into two separate output streams.
	This is also supported by the SDR improvements, which for output stream one ("Speaker 1") is 13.7 dB, and for output stream two ("Speaker 2") is 12.1 dB. 
	%
    %
	%

	\section{Conclusion and discussion}\label{sec:conclusion}
	In this paper, we have introduced the utterance-level Permutation Invariant Training\;(uPIT) technique for speaker independent multi-talker speech separation. 
	We consider uPIT an interesting step towards solving the important cocktail party problem in a real-world setup, where the set of speakers is unknown during the training time.
	
	Our experiments on two- and three-talker mixed speech separation tasks indicate that uPIT can indeed effectively deal with the label permutation problem. These experiments show that bi-directional Long Short-Term Memory\;(LSTM) Recurrent Neural Networks\;(RNNs) perform better than uni-directional LSTMs and Phase Sensitive Masks\;(PSMs) are better training criteria than Amplitude Masks\;(AM). 
	Our results also suggest that the acoustic cues learned by the model are largely speaker and language independent since the models generalize well to unseen speakers and languages. 
	More importantly, our results indicate that uPIT trained models do not require \emph{a priori} knowledge about the number of speakers in the mixture. Specifically, we show that a single model can handle both two-speaker and three-speaker mixtures. This indicates that it might be possible to train a universal speech separation model using speech in various speaker, language and noise conditions.

	The proposed uPIT technique is algorithmically simpler yet performs on par with DPCL \cite{hershey_deep_2016,isik_single-channel_2016} and comparable to DANets\cite{chen_deep_2017,chen_single_2017}, both of which involve separate embedding and clustering stages during inference. 
	Since uPIT, as a training technique, can be easily integrated and combined with other advanced techniques such as complex-domain separation and multi-channel techniques, such as beam-forming, uPIT has great potential for further improvement.

	\section*{Acknowledgment}
	We would like to thank Dr. John Hershey at MERL and Zhuo Chen at Columbia University for sharing the WSJ0-2mix and WSJ0-3mix datasets and for valuable discussions. We also thank Dr. Hakan Erdogan at Microsoft Research for discussions on PSM.

	\ifCLASSOPTIONcaptionsoff
	\newpage
	\fi
	%
	%
	%
	%
	%
	%
	%
	\bibliographystyle{IEEEtran}
	\bibliography{mybib}
	%
	%
	%
	%
	%
	%
	%
	\vspace{-10mm}
	\begin{IEEEbiography}[{\includegraphics[width=1in,height=1.25in,clip,keepaspectratio]{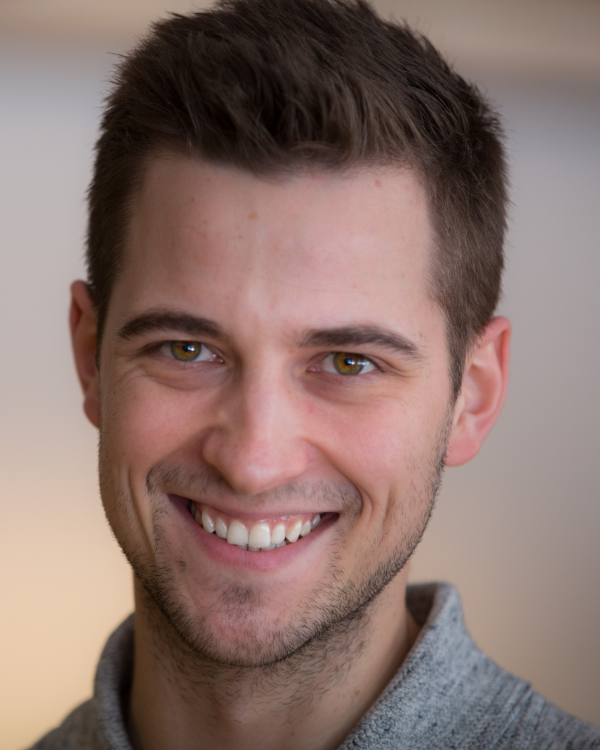}}]%
		{Morten Kolbæk}
		received the B.Eng. degree in electronic design at Aarhus University, Business and Social Sciences, AU Herning, Denmark, in 2013 and the M.Sc. in signal processing and computing from Aalborg University, Denmark, in 2015. 
		He is currently pursuing his PhD degree at the section for Signal and Information Processing at the Department of Electronic Systems, Aalborg University, Denmark.
		His research interests include speech enhancement, deep learning, and intelligibility improvement of noisy speech. 
	\end{IEEEbiography}
	\vspace{-10mm}
	\begin{IEEEbiography}
		[{\includegraphics[width=1in,height=1.25in,clip,keepaspectratio]{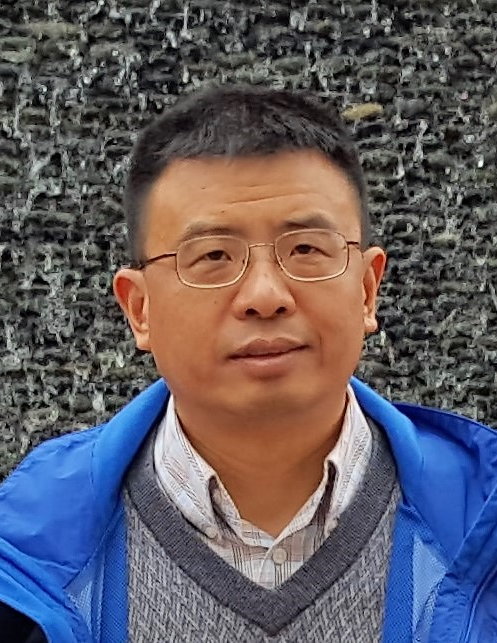}}]
		{Dong Yu}
		(M'97-SM'06) is a distinguished scientist and vice general manager at Tencent AI Lab. Before joining Tencent, he was a principal researcher at Microsoft Research where he joined in 1998. His pioneer works on deep learning based speech recognition have been recognized by the prestigious IEEE Signal Processing Society 2013 and 2016 best paper award. He has served in various technical committees, editorial boards, and conference organization committees.
	\end{IEEEbiography}
	\vspace{-10mm}
	\begin{IEEEbiography}
		[{\includegraphics[width=1in,height=1.25in,clip,keepaspectratio]{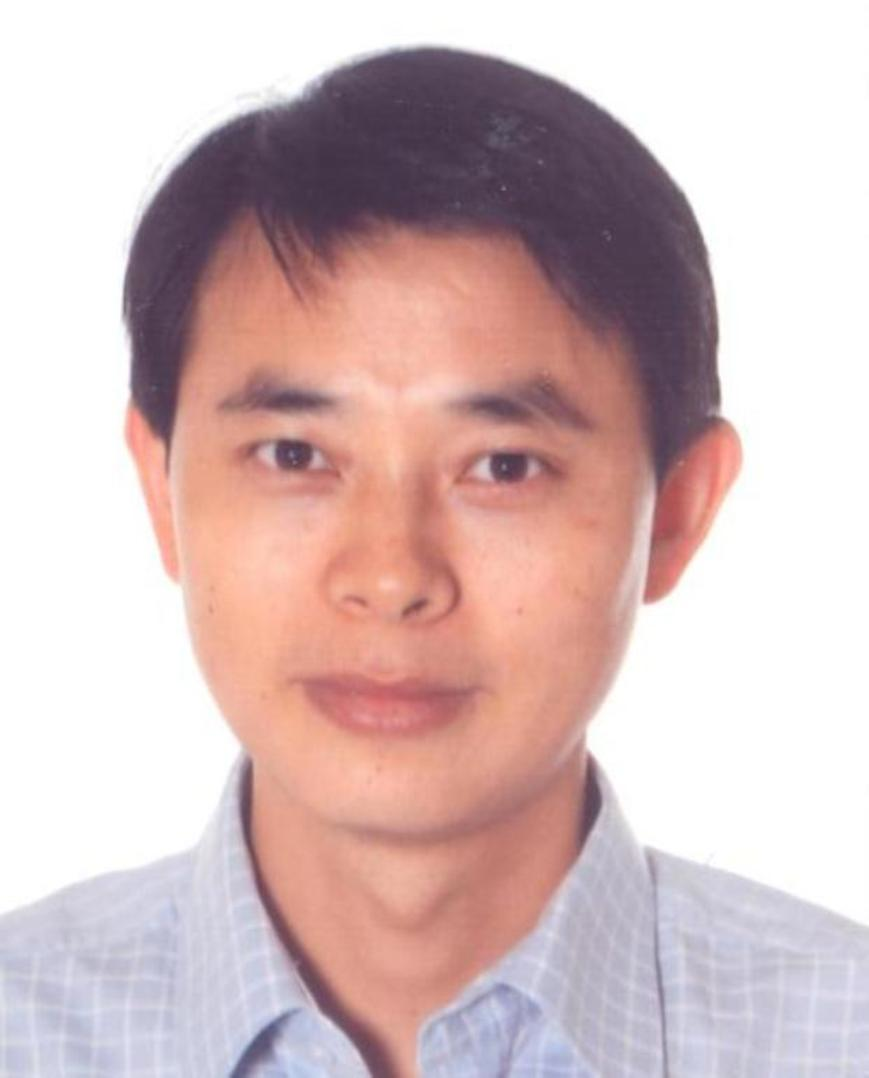}}]
		{Zheng-Hua Tan}
		(M’00–SM’06) is an Associate Professor and a co-head of the Centre for Acoustic Signal Processing Research (CASPR) at Aalborg University, Denmark. He was a Visiting Scientist at MIT, USA, an Associate Professor at SJTU, China, and a postdoctoral fellow at KAIST, Korea. His research interests include speech and speaker recognition, noise-robust speech processing, multimodal signal processing, social robotics, and machine learning. He has served as an Associate/Guest Editor for several journals.
	\end{IEEEbiography}
	\vspace{-10mm}
	\begin{IEEEbiography}[{\includegraphics[width=1in,
			height=1.25in,clip,keepaspectratio]{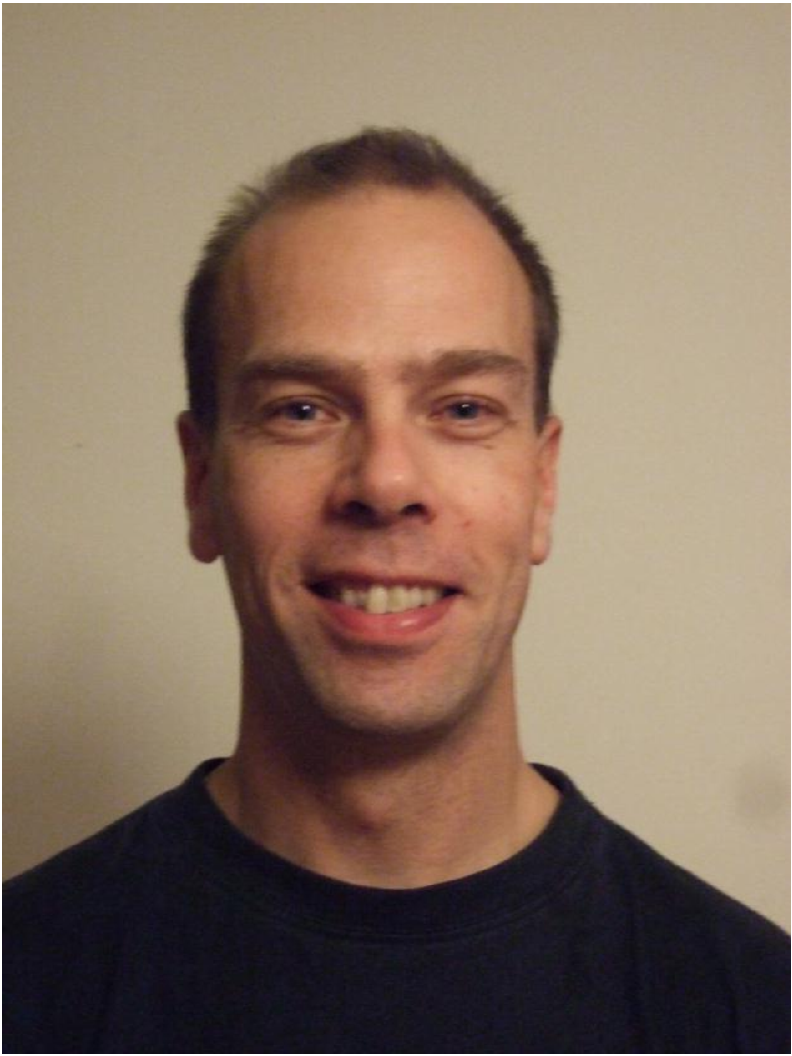}}]{Jesper Jensen}
		is a Senior Researcher with Oticon A/S,
		Denmark, where he is responsible for scouting and
		development of signal processing concepts for hearing instruments. He is also a Professor in Dept.\ Electronic Systems, Aalborg University. 
		He is also a co-head of the Centre for Acoustic Signal Processing Research 
		(CASPR) at Aalborg University. 
		His work on speech intelligibility prediction received the 2017 IEEE Signal Processing Society's best paper award.
		His main interests are in the area of
		acoustic signal processing, including signal retrieval from noisy
		observations,
		intelligibility enhancement of speech signals, signal processing for
		hearing aid applications, and perceptual aspects of signal processing.
	\end{IEEEbiography}
	\vfill\eject
	
	
	

\end{document}